\documentclass{aastex62}
\usepackage{booktabs, url, amsmath}
\usepackage{lineno}
\def\lea{\mathrel{<\kern-1.0em\lower0.9ex\hbox{$\sim$}}}
\def\gea{\mathrel{>\kern-1.0em\lower0.9ex\hbox{$\sim$}}}

\graphicspath{{./}{./Figures/}{figures/}}
\received{}
\revised{}
\accepted{}

\shorttitle{Stellar Clusters in Arp 220 }
\shortauthors{Chandar et al.}

\begin{document}

\title{Arp~220: A Post-starburst Galaxy With Little Current Star Formation Outside of its Nuclear Disks}
\author{Rupali Chandar}
\affiliation{Ritter Astrophysical 
Research Center, University of Toledo, Toledo, OH 43606, USA}
\author{Miranda Caputo}
\affiliation{Ritter Astrophysical Research Center, University of Toledo, Toledo, OH 43606, USA}
\author{Sean Linden}
\affiliation{Dept. of Astronomy, University of Massachusetts at Amherst, Amherst, MA 01003, USA}
\author{Angus Mok}
\affiliation{OCAD University, Toronto, Ontario, M5T 1W1, Canada}
\author{Bradley C. Whitmore}
\affiliation{Space Telescope Science Institute, 3700 San Martin Drive, Baltimore, MD, 21218, USA}
\author{Daniela Calzetti}
\affiliation{Dept. of Astronomy, University of Massachusetts at Amherst, Amherst, MA 01003, USA}
\author{Debra M. Elmegreen}
\affiliation{Department of Physics \& Astronomy, Vassar College, Poughkeepsie, NY 12604, USA}
\author{Janice C. Lee}
\affiliation{Gemini Observatory/NSF’s NOIRLab, 950 N. Cherry Avenue, Tucson, AZ, 85719, USA}
\author{Leonardo Ubeda}
\affiliation{Space Telescope Science Institute, 3700 San Martin Drive, Baltimore, MD, 21218, USA}
\author{Richard White}
\affiliation{Space Telescope Science Institute, 3700 San Martin Drive, Baltimore, MD, 21218, USA}
\author{David O. Cook}
\affiliation{Caltech/IPAC, 1200 E. California Boulevard, Pasadena, CA 91125, USA}
\correspondingauthor{Rupali Chandar}
\email{Rupali.Chandar@utoledo.edo}

\begin{abstract}
The ultra-luminous infrared galaxy Arp~220 is a late-stage merger with several tidal structures in the outskirts and two very compact, dusty nuclei that show evidence for extreme star formation and host at least one AGN.  
New and archival  high-resolution images taken by the {\em Hubble Space Telescope} provide a state-of-the-art view of the structures, dust, and stellar clusters in Arp~220.  These images cover the near-ultraviolet, optical, and near-infrared in both broad- and narrow-band filters. We find that $\sim90$\% of the H$\alpha$ emission arises from a shock-ionized bubble emanating from the AGN in the western nucleus, while the nuclear disks dominate the P$\beta$ emission.  Four very young ($\sim3-6$~Myr) but lower mass ($\lea 10^4~M_{\odot}$) clusters are detected in H$\alpha$ within a few arcsec of the nuclei, but produce less than $1$\% of the line emission.
We see little evidence for a population of massive clusters younger than 100~Myr anywhere in Arp~220, unlike previous reports in the literature.
From the masses and ages of the detected clusters, 
we find that star formation took place more-or-less continuously starting approximately a $\mbox{few}$~Gyr ago with a moderate rate between $\approx3-12~M_{\odot}~\mbox{yr}^{-1}$. 
Approximately 100~Myr ago, star formation shut off suddenly everywhere (possibly due to a merging event), except in the nuclear disks.  A very recent flicker of weak star formation produced the four young, low-mass clusters, while the rest of the galaxy appears to have remained in a post-starburst state.  
Cluster ages indicate that the tidal structures on the west side of the galaxy are older than those on the east side, but all appear to pre-date the shutoff of star formation.
Arp~220 has many of the characteristics expected of a 'Shocked Post-Starburst Galaxy' or SPOG, since most of the system has been in a post-starburst state for the past $\sim100$~Myr and the detected H$\alpha$ emission arises from shocked rather than photo-ionized gas.
\end{abstract}
\section{Introduction} \label{sec:intro}
\par
At a distance of  $\sim88$ Mpc \citep[e.g.][]{Parra07,Armus09}, Arp~220 is the closest Ultra-Luminous Infrared Galaxy (ULIRG), with $\log L_{\rm IR} / L_\odot = 12.21$ \citep{Sanders03}  and a stellar mass of log~$M_*\approx 10.8$ \citep{Vivian12}. This late-stage merger contains two compact ($\approx150$~pc), highly obscured nuclei (east and west) separated by $\approx 400 {\rm ~pc~} (1'')$ which dominate the infrared luminosity.  There is clear evidence for an AGN in the western, more infrared-luminous nucleus, and partial but inconclusive evidence for an AGN in the eastern nucleus.  Long-term monitoring with VLBI has revealed around 100 point sources, believed to be supernovae (SNe) and supernova remnants (SNR) resulting from the extreme star formation occurring in the compact nuclei \citep[e.g.][]{Varenius19}.  Outside of the nuclear disks, the central $12.5\arcsec$ shows significant obscuration by dust in some locations, and there are several faint but distinct tidal features in the outskirts on both the west and the east sides of the system.

Ages of the stellar populations in the different structures within Arp~220 provide important clues to its star formation and merger/interaction history.  \citet{Wilson06} concluded that the system has formed a rich population of very young ($<10$ Myr), extremely massive ($\sim10^6 - 10^7~M_\odot$) clusters, plus a less-dominant intermediate age (70-500 Myr) population with lower masses $\sim10^5 - 10^6~M_\odot$ located at larger radii, based on somewhat limited $HST$ imaging of the central region. Spectroscopy of the diffuse starlight in the outer portions of the system suggests these regions are dominated by an intermediate-age stellar population ($\sim500-900$~Myr),  with a young stellar population ($<100$ Myr) that has an increasing contribution as one moves towards the center of the galaxy \citep{Zaurin08}. The older population has ages similar to those found for stars in many post-starburst galaxies \citep[e.g.][]{Dressler83,Couch87,French18}. 

\begin{figure}[!ht]
	\centering
\includegraphics[width=18.5cm]{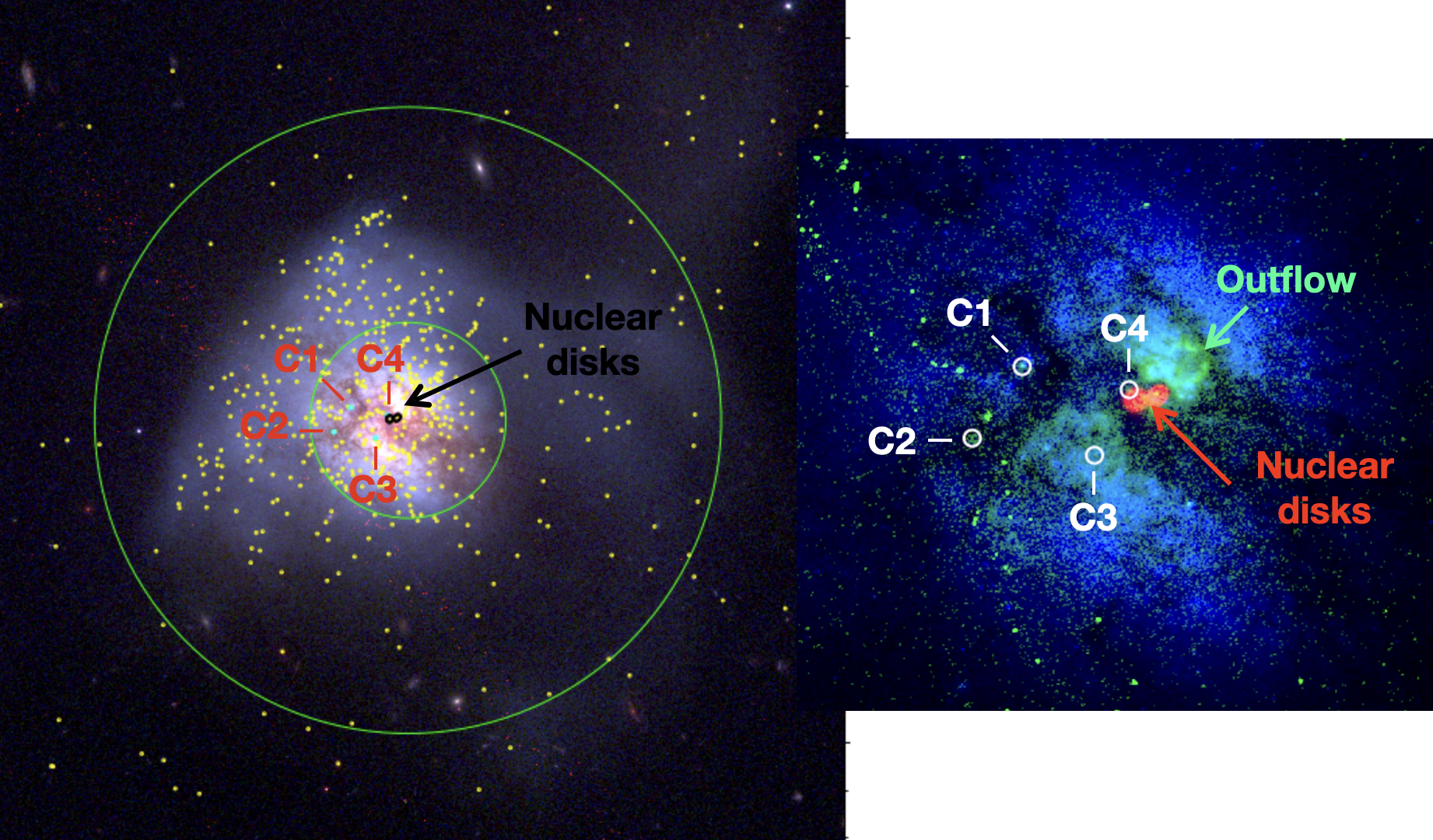}
	\caption{{\bf Left:} An $HST$ $BVI$ image of Arp 220 identifies the locations of the 2 nuclear disks (small black circles), the inner $12.5\arcsec$ ($\approx5$~kpc) and outer region beyond $40\arcsec$.  Clusters, shown as yellow dots, are detected throughout the system, including very close to the dusty nuclear disks.  The strip of red dots running diagonally through this image are artifacts resulting from the gap between detectors in the ACS camera.
 {\bf Right:} Three color image of the central portion of Arp~220 created from the B-band (blue), H$\alpha$ (green), and 6~GHz (red) emission. The nuclear disks and shocked, outflowing H$\alpha$ gas are identified, along with 4 very young ($<10$~Myr) line-emitting clusters identified in MUSE observations \citep{Perna20}. The bright green points which fall diagonally from the top-left side of the image to the bottom-middle in the right panel (also seen in the upper right panel of Figure~2) are artifacts from cosmic-ray hits.
 The images are oriented with North up and East to the left.
	}\label{fig:arp220}
\end{figure}

The central $12.5\arcsec$ ($\approx5$~kpc) of Arp~220 is criss-crossed by dust, with significant obscuration in places, as shown in Figure~\ref{fig:arp220}.  Because of this patchiness, the range of ages of the stellar populations in the central $12.5\arcsec$ but outside of the nuclear disks are still not well known.  In this work, we use an age-dating method that allows the maximum cluster reddening to vary during the fitting process based on the amount of dust in the region.  The results from this method allow us to directly address: 
{\em How much star formation has taken place in the last 10~Myr outside of the nuclear disks ? How much star formation took place between 10 and 100~Myr ago ?  Was star formation over the past Gyr bursty or continuous? Do different regions and structures within Arp~220 have stellar populations with different ages ? } The answers to these questions provide critical clues to the star formation and merger/interaction history of Arp~220, and are the focus of this paper.

\par
A direct and independent method of measuring the recent star formation history of galaxies over at least the past 0.5~Gyr is to use the ages and masses of their stellar clusters. 
Young clusters directly trace the star formation process, since their numbers and masses scale with the rate of star formation \citep[e.g.,][]{Larsen02,Chandar17,Whitmore20}, although some authors report a more complicated relationship  \citep[e.g.,][]{Bastian12,Adamo15,Krumholz19}.

In this work, we revisit Arp~220 and it's cluster population by taking advantage of new, high resolution observations that cover the entirety of this system from the near-ultraviolet through the near-infrared.
The observations were taken as part of the Clusters, Clumps, Dust, and Gas in Extreme Star-Forming Galaxies (CCDG) survey (PI: Chandar), which has recently collected both broad- and narrow-band imaging of 13 of the most extreme, star-forming galaxies found in the nearby (D $< 100$ Mpc) Universe. Our observations significantly extend the \citet{Wilson06} coverage to the entire optically bright portion of Arp 220, and provide the definitive view of the cluster population in Arp~220. 

\par

The rest of this paper is organized as follows. Section 2 presents multi-wavelength (near ultraviolet through near infrared) $HST$ images of Arp~220, cluster selection, and photometry. We examine key features in Arp~220, including the central dusty region and tidal tails using the measured colors of clusters in Section~3. In Section~4 we constrain the reddening and use it to derive the ages and masses of the clusters; Section 5 presents the cluster mass function and determines the shape at the upper end.  In Section~6 we estimate the star formation rate in different intervals of age, and in Section~7 we discuss these results in the context of galaxy formation and evolution.  Our main results are summarized in Section~8.
\section{Hubble's View of Arp~220} \label{sec:obs}
\subsection{Observations}
\begin{figure}[!ht]
	\centering
\includegraphics[width=18.5cm]{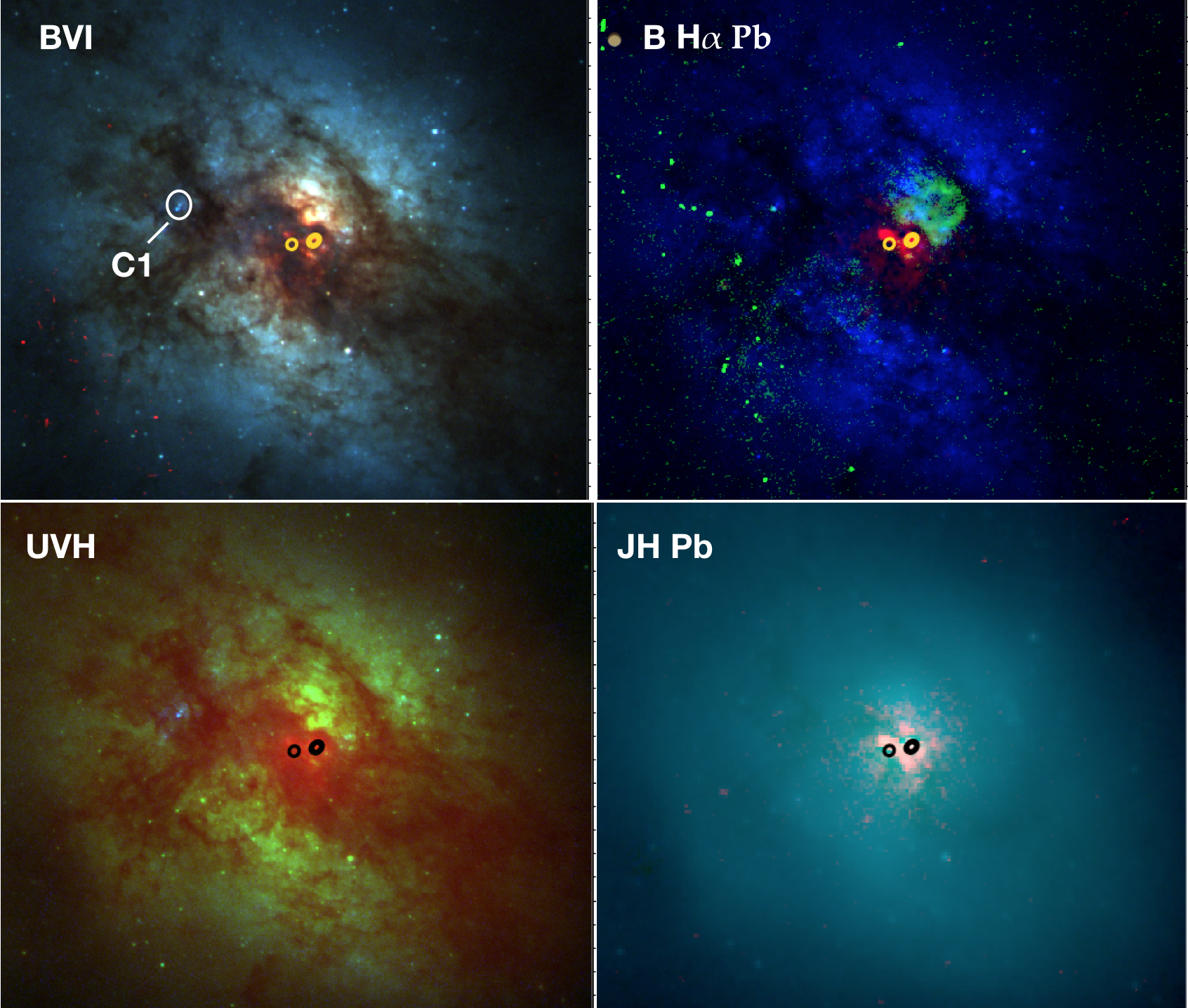}
	\caption{We show four different filter combinations with blue-green-red as indicated in each panel, and which highlight different features in the central portion of Arp~220.  The top-left panel ($BVI$) highlights the dust lanes.  One of two young $<10$~Myr clusters detected in our $HST$ images is identified.  The top-right panel shows that H$\alpha$ and Paschen $\beta$ emission are not co-spatial; the horseshoe shaped H$\alpha$ emission in green is produced by outflows from West nuclear disks which hosts an AGN, while most of the Pa $\beta$ emission comes from the nuclear disks and the area immediately surrounding them.  The bottom-left panel shows a wide-wavelength coverage of broad-band filters and the bottom-right panel highlights new, NIR images in J and H band.  The small yellow and black circles mark the locations of the 2 nuclear disks.}\label{fig:Arp220color}
\end{figure}
\par
The {\em Hubble Space Telescope} ({\em HST}) has imaged Arp~220 in broad- and narrow-band filters covering the near-ultraviolet to the near-infrared.  
  The observations used in this work are a mix of new and archival data. New observations with the WFC3 camera were taken as part of our program GO-15649 (PI:Chandar) in the NUV (F275W), U (F336W), V (F555W), P$\beta$ (F130N), and H (F160W) bands, and archival images are available in the B (F438W), I (F814W), and H$\alpha$ (F665N) bands  (the first two taken with the ACS camera, and the latter with the WFC3).
Figure~\ref{fig:arp220} shows an image of Arp 220 which captures the multiple extended tail-like features in the outer regions as well as the bright, dusty central region. The faint, diffuse emission in the tidal features is dotted with point sources, which are stellar clusters.  The central region is criss-crossed by dark dust lanes.  Despite the dust, bright starlight and point-like clusters are clearly seen throughout the central region.  We follow \citet{Scoville17} and adopt a luminosity distance of 88~Mpc to Arp~220 \citep{Armus09}, which gives a scale where $1\arcsec \approx 400$~pc.  
\par 
Individual exposures are processed through the standard Pyraf/\linebreak[0]{}STSDAS CALACS or CALWFC3 software, which performs initial data quality flagging, bias subtraction, gain correction, bias stripe removal, correction for CTE losses, dark current subtraction, flat-fielding and photometric calibration.
These individual files are aligned and drizzled onto a common grid to create a single image per filter using the DRIZZLEPAC software package.  The V band image is used as the reference, with a pixel scale of $0\farcs04$ (the native WFC3 pixel scale), and astrometry calibrated with GAIA DR2 sources \citep{gaia2, gaia2astrometry}.
Sky subtraction is performed during the drizzling process and the output images are all oriented with North up and East to the left.  The final FITS files are in units of
electrons per second.
\par 
We create a  continuum-subtracted P$\beta$ line map for Arp~220 by first using the F110W and F160W broad-band WFC3/IR images to estimate the stellar continuum at 1.28 $\mu$m ($F_{1.28\mu m}$) in each pixel. Several bright isolated stars, which do not have associated recombination line emission, are used to determine the scale factor needed to remove the stellar contribution from the narrow-band filter.  We find: $F_{P_{\beta}} = F_{130N} - a \times F_{1.28\mu m}$, with a range of $a$ values between 0.98 and 1.03, and adopt a best-fit scale factor of $a=1.015$.  We use a similar technique to create a continuum-subtracted H$\alpha$ image, but use the V and I filters to estimate the stellar continuum.  While the P$\beta$ image has a significantly longer exposure time than H$\alpha$ (2000~sec vs. 400~sec), the H$\alpha$ images has higher resolution, better for identifying clusters. The expected H$\alpha / \mbox{P}\beta$ ratio for an HII region with no extinction (assuming case B recombination) is $\sim17$.  For both of these reasons, we expect the H$\alpha$ image to better identify all but the most heavily extincted line-emitting clusters.
\par 
In this study, we define three regions within Arp~220, shown in Figure~\ref{fig:arp220}: (1) the two small, nuclear disks, (2) the inner $12.5\arcsec$ ($\sim5$~kpc) region demarcated by the inner circle, which is dominated by the bulge and by the presence of dust, and (3) an outer region which has little dust and includes several tidal features. These regions are discussed in detail in Section~3. 

\subsection{Color Images of the Inner Region}

In Figure~\ref{fig:Arp220color}, we show four different filter combinations of the inner region of Arp~220.  These composite images reveal different structures and provide insight to the cluster population and dust in this region. 

\begin{itemize}

    \item The $BVI$ figure in the top-left panel of Figure~2 gives a good view of the dust criss-crossing the central region of Arp~220.  It also shows that a significant portion ($\approx50$\%) of the central $12.5\arcsec$ is {\em not} strongly affected by dust.  Many point-like clusters are seen in the optically bright and the moderately dusty regions.  The red streaks in the lower-left portion of this figure are caused by artifacts in the $I$ band (due to too few exposures).

\item  The $B$ H$\alpha$ P$\beta$ image in the top-right panel of Figure~2 highlights the contribution and location of line emission.  One striking result is that H$\alpha$ and P$\beta$ emission originate from completely different regions in Arp~220. The dominant H$\alpha$ (green) feature is the shocked, outflowing 'superbubble' originating from the AGN in the west nuclear disk \citep{Lockhart15}. 
The most intense P$\beta$ emission arises from the two nuclear disks, which are embedded in extended, diffuse P$\beta$.

\item  The $UVH$ image in the bottom-left panel emphasizes faint, blue cluster C1, and does not reveal any obvious, red clusters in the dust lanes.

\item  The $JH$ P$\beta$ image in the lower-right panel is the best filter combination currently available to identify very young, still embedded clusters which are obscured at optical wavelengths. We have carefully searched this image and find a dozen broad-band sources in dust lanes; none however, have P$\beta$ line emission. Several of these are visible in the I band (but not at shorter wavelengths).  We discuss the nature of these clusters in Section~3.2.

\footnote{While unsharp-masked 8$\mu$m images with Spitzer often reveal young, embedded clusters, this band is saturated in observations of Arp~220, presumably due to the AGN.}

\end{itemize}

In Figure~\ref{fig:arp220}, we identify the only (four) line-emitting clusters identified thus far in Arp~220, designated C1, C2, C3, and C4.  These were discovered in MUSE optical IFU spectra \citep{Perna20}, and help guide searches for additional very young, line emitting clusters.
Source C1 is the brightest blue cluster observed in the $HST$ images.  We measure an H$\alpha$ flux of $\approx2.5\times10^{-18}~\mbox{erg~s}^{-1}$ and P$\beta \approx 3.5\times10^{-19}~\mbox{erg~s}^{-1}$ in a 5 pixel aperture, consistent with the modest level of reddening determined by \citet{Perna20} for this cluster. C3 is at the limit of what we can detect in the line maps, and has measured H$\alpha \approx P\beta \approx 5\times 10^{-19}~\mbox{erg~s}^{-1}$.
Clusters C1 and C2 are also visible in the broad-band images.
C3 and C4 are fainter and not detected in our $V$-band image (see below). 

\subsection{Cluster Selection and Photometry}

At the distance of Arp~220, clusters appear as point-like sources while individual stars are not detectable.  
The cluster catalog was created by running the IRAF/DAOFIND task on the $V$ band image with a signal-to-noise threshold of 3 above the background.  This procedure detected almost all obvious sources with very few spurious detections. 
Aperture photometry was performed in a two-pixel radius for all sources in the five broad-band NUV through optical filters, with the background level determined in an annulus with radii between 7 and 9~pixels around each source.  The following zero points from the on-line instrument handbook were applied to convert the apparent magnitudes to the VEGAMAG system: 22.640 (WFC3/F275W), 23.526 (WFC3/F336W), 25.764 (ACS/F438W) , 25.832 (WFC3/F555W), and 25.518 (WFC3/F814W).
Filter-dependent aperture corrections  of 1.097~mag (F275W), 0.991 (F336W), 0.821 (F435W), 0.741 (F555W), and 0.913 (F814W) were determined from isolated sources and applied to obtain the total apparent magnitude for each cluster. While our photometry is in the VEGAMAG system, we refer to the filters as NUV (F275W), U (F336W), B (F438W), V (F555W), and I (F814W) throughout.
Our final catalog includes 596 cluster candidates down to a magnitude limit of $m_V=27$~mag ($M_V\sim-8$). 

Clusters are detected throughout Arp~220, including in all tidal features. It is important to note that many optically bright clusters are detected in the central $12.5\arcsec$, including within $\approx1~\arcsec$ of the extremely dusty region where the nuclear disks are located.  Nearly 40\% (223) of the 596 clusters in our sample are found within the inner $12.5\arcsec$.

Crowding of recently formed clusters is one of the most challenging issue related to cluster detection and photometry. Luckily, the cluster population in Arp~220 is significantly more dispersed than in well-studied on-going merger systems like the Antennae and NGC~3256.  Although these galaxies are 2 and 4 times closer than Arp~220 respectively, their strong star and cluster formation makes cluster detection more challenging than in Arp~220.

\section{Regions of Interest}

In this Section, we focus on key features found in three distinct regions within Arp~220: (1) the nuclear disks, (2) the central $12.5\arcsec$ outside of these disks, and (3) several tidal features located towards the outskirts of the system.  We examine the broad distribution of cluster colors in different regions within Arp~220 (leaving detailed age dating to Section~4), but not in the nuclear disks, which are too dusty and compact to identify individual clusters even at HST resolution.

\subsection{The Nuclear Disks of Arp~220}

As highlighted in Figure~1, the center of Arp~220 contains two very compact ($\approx150$~pc) nuclear disks. Despite covering less than 0.1\% of the area of the galaxy, most of the current star formation appears to be taking place in these disks. They are heavily obscured at optical wavelengths, but emit strongly in Pa$\beta$. Very high resolution radio, sub-mm, and infrared imaging has revealed $\approx100$ total point sources in the two disks, almost certainly supernovae and supernova remnants, the signature of recent, intense star-formation (e.g., Smith et al. 1998; Varenius et al. 2018).  

There is clear evidence for an embedded AGN in the western nucleus of Arp~220 \citep[e.g][]{Rangwala15,Lockhart15,Barcos-Munoz18}. A number of observations, including the detection of Fe K$\alpha$ emission at 6.3 keV support this conclusion \citep[e.g.][]{Yoast-Hull17,Wang21}. \citet{Varenius16} further suggested that the alignment of the 150 MHz continuum morphology above and below the nuclear disk, and the similarity to the 150 MHz observations of M82, are evidence for an outflow in the eastern nucleus, with the southern side being the closer (approaching) part consistent with CO (6-5) observations presented in \citet{Rangwala11}. This suggests the eastern nucleus may also contain an AGN, although this is not yet certain.

\subsection{Inner $12.5\arcsec$ (Outside of Nuclear Disks)}
\begin{figure}[!ht]
	\centering
\includegraphics[width=14.0cm]{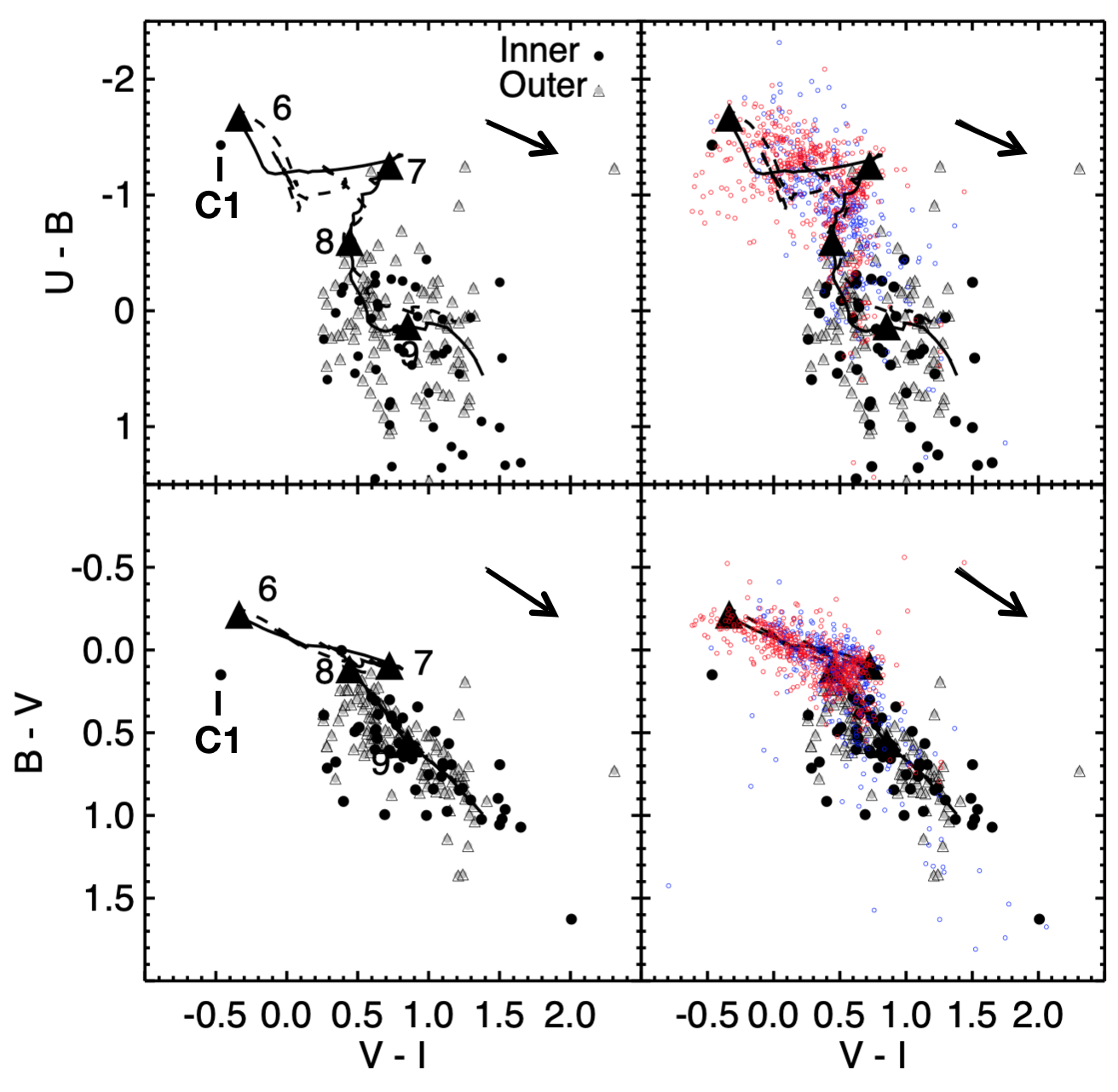}
	\caption{The panels on the left show the U-B vs. V-I (top-left) and B-V vs. V-I (bottom-left) color-color diagrams for clusters detected in Arp~220; the photometry is in the VEGAMAG system but for ease we refer to the filters by their Johnson names, as described in Section~2.3. Predictions from \citet{Bruzual03} show the expected aging of clusters from soon after their birth around log~$(\tau/\mbox{yr})=6$ to the ages of globular clusters for solar metallicity (solid line) and $1/4\times$solar (dashed line).  
	We plot clusters brighter than $m_V \leq 25.5$~mag found inside (outside) of $12.5\arcsec$ as circles (triangles).  The one young, blue cluster seen close to the youngest portion of the model track is source C1.  There are essentially no clusters with colors between the log~$(\tau/\mbox{yr})=6 - 8$ model tracks.  The right panels show that the colors of the cluster population in Arp~220 is quite different from those in the on-going mergers in the Antennae (red) and NGC~3256 (blue).  
	}\label{fig:twocolor}
\end{figure}

Although this region contains much of the dust that Arp~220 is famous for, a large number of clusters are clearly visible in the optical and NUV images.
In Figure~\ref{fig:twocolor} we show the measured $U-B$ vs. $V-I$ colors of clusters brighter than $m_V$ of $25.5$~mag (the $U$ band photometry becomes less certain for sources fainter than this).  Clusters in the inner $12.5\arcsec$ portion of Arp~220 are shown as black circles and those between $12.5$ and $45\arcsec$ as gray triangles. The cluster colors are compared with predictions from the stellar evolution models of \citet{Bruzual03} (hereafter BC03), which closely match the observed colors of clusters in nearby spiral galaxies (Lee et al. 2021; Turner et al. 2021). The models predict the luminosity and color evolution of clusters starting soon after they form (ages $\sim1$~Myr), through the ages of ancient globular clusters of $\approx12$~Gyr. The arrow in each panel shows the direction of reddening, assuming a Galactic extinction low (Fitzpatrick 1999). We show the solar metallicity BC03 model with no nebular contribution.

\begin{itemize}
\item A single cluster, labelled C1 in Figures~\ref{fig:arp220} and \ref{fig:Arp220color}, has (U-B) vs. (V-I) colors that indicate an age of just a few Myr. It falls slightly left of the models in the region expected of clusters with H$\alpha$ emission.  
Source C2 is also in our V-band selected sample, has fairly blue colors, but is fainter than $m_V=25.5$~mag and therefore not shown in the figure. 
All other clusters detected in our $HST$ images are significantly redder than C1 and C2.  We inspected our P$\beta$ continuum-subtracted image and found no line emission from any cluster besides C1$-$C4.

\item As one moves along the cluster model track starting from the youngest ages at the upper-left, the colors of clusters start abruptly around $\approx10^8~$yr, and  continue red-ward more-or-less continuously along the model towards older ages of at least several Gyr.

\item Clusters located within $12.5\arcsec$ (solid circles) have somewhat redder $U-B$ colors than those located further than this (gray triangles), including a number which fall off the models to the red side.  This is almost certainly due to reddening by dust in the central region.

\item The B-V vs. V-I two-color diagram in the lower-left has less scatter around the models than U-B vs. V-I (upper-left) and supports the points made above from the U-B vs. V-I diagram. 
 
\end{itemize}

Two-color diagrams give important constraints on the ages and reddening of the clusters.  Almost all clusters have colors indicating they are at least $100$~Myr old. We have carefully inspected our continuum-subtracted H$\alpha$ and Pa$\beta$ images, and find that none of the clusters in our catalog have any detectable Hydrogen line emission, except C1-C4, consistent with the ages indicated by the broad-band colors. If other very young ($\tau \lea 6$~Myr), moderately reddened clusters were present, we would detect their H$\alpha$ emission.

We estimate that $\sim90$\% of the H$\alpha$ line emission (measured from the continuum subtracted $HST$~H$\alpha$ image) comes from the AGN-driven shocked gas bubble rather than from recently formed clusters.  Therefore, estimates of the SFR in Arp~220 which rely on global H$\alpha$ intensity should be viewed with caution. 
\par 
In the right panels of Figure~\ref{fig:twocolor}, we compare the colors for clusters in Arp~220 with those in the Antennae (red) and NGC~3256 (blue), two ongoing merging systems. Both systems are dominated by very young, blue clusters, which is quite different from Arp 220.  Although not shown, the distribution of cluster colors in the Antennae and NGC~3256 are similar to those of cluster populations in spirals, irregulars and dwarf starbursts.
The color-color diagrams of optically-detected clusters indicate that, outside of the nuclear disks, Arp~220 has nearly stopped forming clusters altogether over the past 100~Myr. 
While the Arp~220 cluster color distribution is quite different from that observed in {\em spirals, irregulars, and mergers}, it is similar to the U-B vs. B-R color-color diagram for the cluster population in the {\em post-starburst} galaxy S12 \citep{Chandar21}. In Section~4.2 we assess the possibility that Arp~220 has formed young, massive clusters that remain embedded or obscured by dust, after estimating the amount of reddening in Section~4.1.

\subsection{Outer Region and Tidal Features}
\begin{figure}[!ht]
	\centering
	\includegraphics[width=16cm]{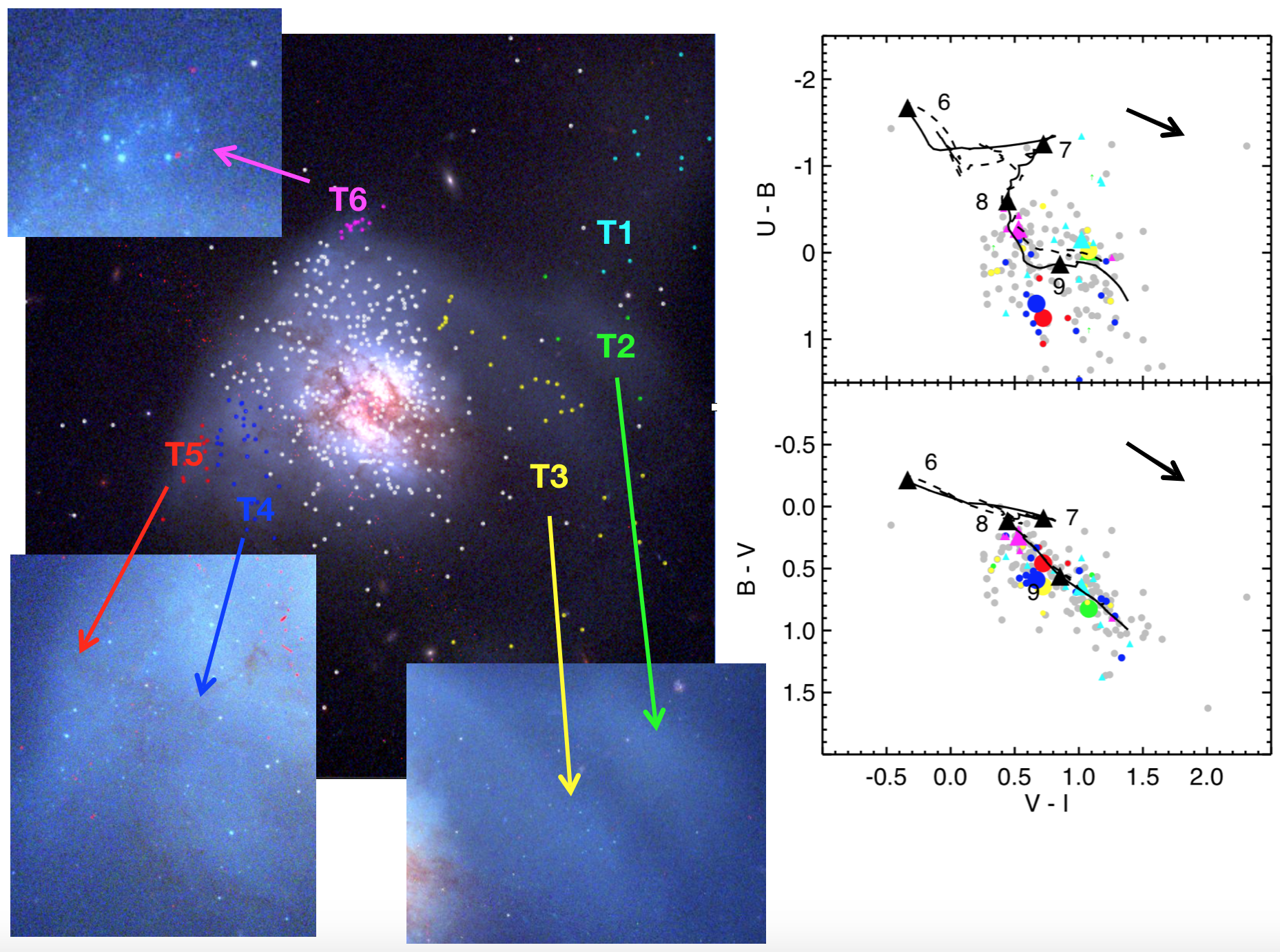}
	\caption{{\bf Left:} Six tidal features in Arp~220 and their associated clusters are identified in the $HST$ BVI color image.  These are discussed in Section~3.3. {\bf Right:} The U-B vs. V-I (top) and B-V vs. V-I (bottom) colors of clusters in the six features are plotted.  The median cluster colors are shown as the larger symbols, and used to constrain the ages of the stellar populations in each tidal feature.  See text for details.  }\label{fig-tidal}
\end{figure}
\par
\begin{table}[ht]
	\caption{Tidal Feature Properties}\label{tab:TT}
	\centering
	\begin{tabular}{lcccccc}
	Tidal & Median & Median & Median & Typical  & Approximate \\ 
	Feature & $U-B$ & $B-V$ & $V-I$ & Age & Length (kpc) &  \\ \hline \hline
T1 & 0.0 & 0.7 & 1.0 & $1-2\times10^9$ & 19   \\
T2 & 0.12  & 0.8 & 1.05 & $1-2\times10^9$ & 20   \\ 
T3 & 0.1 & 0.7  & 0.7 & $1-2\times10^9$ & 32  \\
T4 & 0.6 &  0.6 & 0.65 & $6\pm1\times10^{8}$ &  7.9  \\
T5 & 0.7 & 0.48  & 0.7 & $6\pm1\times10^{8}$ & 8.2   \\
T6 & -0.1 & 0.3  & 0.5 & $1.5-2\times10^8$ &  3.5  \\
        \hline
	\end{tabular}
	\end{table}
Tidal tails form during interactions between galaxies.  The ages of the stellar populations in these tails provide clues to the formation history of the system,  and are particularly interesting for Arp~220, which is an advanced merger.
In Figure~\ref{fig-tidal}-left, we identify 6 distinct tidal features in Arp~220  from their diffuse emission.  All are located outside of $12.5\arcsec$.  T1 (cyan), T2 (green), and T3 (yellow) are on the west side of the galaxy, and fainter than tidal features T4 (blue), T5 (red), and T6 (magenta) identified on the east side of the galaxy.  
T1 is the faintest tidal structure, 
T3 is long and curved, wrapping nearly a quarter of the way around Arp~220. T1 is oriented nearly orthogonal to T2 and T3, and regions of little diffuse emission separate T2 from T3, and T4 from T5.  We highlight some of the clusters found in the tidal tails of Arp~220 in the three zoomed panels.
\par 
In Figure~\ref{fig-tidal}-right we show the colors of the clusters identified in each tidal structure, using the same color scheme as in the left panel.  The $U-B$ vs. $V-I$ color-color diagram is at the top, and the $B-V$ vs. $V-I$ diagram is below. While the $U-B$ vs. $V-I$ diagram is usually preferred because age/reddening/metallicity are better separated, many of the clusters in the tidal features are quite faint and have large photometric uncertainties in the $U$ band. The outer portion of Arp~220 shows little evidence for dust, so the $B-V$ vs. $V-I$ diagram should give important insight into the relative ages of the tidal features.
The BC03 models are shown for solar (solid line) and $1/4\times$solar metallicity (dashed line), and black triangles track the evolution of (logarithmic) ages from log~$(\tau/\mbox{yr})=6$ in the top left, down to log~$(\tau/\mbox{yr})=9$ towards the lower-right of each panel.  The black arrows show the direction of reddening based on a Milky Way extinction law \citep{Fitzpatrick99}, although we do not believe reddening has much impact on the clusters studied in this section.
\par 
Figure~\ref{fig-tidal}-right shows the colors measured for individual clusters in each tidal feature (small colored circles), as well as the median colors (larger colored circles), using the same color coding as in the left panel.  
We use a magnitude cut of $m_V \lea 25.5$~mag for all tails except for T1 and T2, where we use $m_V \lea 26.5$~mag because most clusters are quite faint.
We see that the clusters in T6 (shown in magenta) have a small spread in colors, as do most clusters in T4, which are shown in blue (especially in the lower panel).  The clusters in the other tails have a larger range of colors.
The median cluster colors from the six different tidal features form a nice sequence along the tracks, particularly in the $B-V$ vs. $V-I$ diagram, and are compiled in Table~\ref{tab:TT}.
  The oldest/reddest median cluster colors are found in tails T1 and T2, whereas those in T4, and T5 are fairly similar and fall between T6 and T1/T2/T3.
\par 
We can use the median cluster colors in each tidal feature to estimate the age of each structure by assuming there is no reddening and reading off the closest ages from the BC03 model.  
T6 (magenta), the most compact and bluest feature, has median colors consistent with an age of $\approx1.5-2\times10^8$~Myr. 
The colors of clusters in T1 (cyan) through T5 (red) are clearly redder than those in T6, so the stellar populations must be older, consistent with Arp~220 being an older merger of several$\times100$~Myr; \citep{Mundell01,Konig12}.  While, it is somewhat challenging to precisely estimate the ages of these features because the age-metallicity degeneracy affects the cluster colors, the median $B-V$ and $V-I$ colors for T4-T5 indicate ages somewhere between $6-8\times10^8$~yr, assuming solar metallicity, while cluster colors in T1-T2-T3 give ages of $\approx 1-2\times10^9$~yr.  We compile these estimated ages, which are similar from both the U-B vs. V-I and the B-V vs. V-I diagrams, in Table~\ref{tab:TT}.  Tails T1 through T5 all have cluster populations that are {\em older} than the youngest, most massive clusters that we see in the main body of Arp~220, indicating they formed before cluster formation ceased in Arp~220. Feature T6 has clusters with ages which are close to, but slightly older than, the time when star formation ceased in the main body of Arp~220.

\par 

Our age constraints based on cluster colors are consistent with those estimated from ground-based spectroscopy of portions of T2 and T3.  \citet{Rodriguez08}    
performed a direct fit of the overall continuum shape in their spectrum using CONFIT, a $\chi^2 $ minimum technique developed by \citet{Tadhunter05}  which fits the  minimum number of stellar components required to model the observations. They find best fit ages between $6-9\times10^8$~yr for the stellar populations in T2 and T3, similar to the constraints we find from the median cluster colors in each tail.
\par 
We make a very approximate estimate of the length of each tidal feature, and compile these measurements (in kpc) in the last column of Table~\ref{tab:TT}.  We find a reasonable correlation between the length of a tidal feature and it's estimated age, such that longer tidal features are older. This is expected if the stellar component of tidal tails spreads out over time.  

\section{Properties of Clusters in Arp~220}

The ages and masses of stellar clusters provide key windows into their lifecycles and into the evolution of their host galaxy.  In this section we estimate the age and mass of the clusters in our Arp~220 sample after constraining the reddening in the dusty central region. We also compare our age-mass results with those made by \citet{Wilson06} for a sub-sample of clusters.

\subsection{Constraints on Reddening}

The biggest challenge to correctly age dating clusters in Arp~220 lies in navigating the well-known age-reddening degeneracy.  This degeneracy arises because redder colors can either indicate that a cluster is older {\em or} that it is affected by dust.
In this section, we estimate the amount of reddening that affects clusters in the inner $12.5\arcsec$ of Arp~220, since dust does not appear to have much impact on the colors of clusters outside this region.
\par 
Several clues help us constrain the ages of the clusters, which in turn helps constrain the amount of reddening.
One of the most important and straightforward clues is the presence or absence of hydrogen emission lines.  Massive ($\gea 10^4~M_{\odot}$) clusters younger than $\approx6$~Myr and moderate amounts of reddening E(B-V)$\lea 3$~mag, $A_V \lea 9$~mag will be detectable in H$\alpha$ (and sometimes P$\beta$; see also \citet{Whitmore11}).
We only find line emission for optically detected clusters C1-C4, but for no other sources. This lack of Hydrogen emission lines suggests that nearly all bright clusters in the central portion of Arp~220 are older than $6$~Myr. In fact, the colors of clusters located furthest away from dust features but still within the central $12.5\arcsec$ match the $\approx100-300$~Myr BC03 models, consistent with age estimates made from integrated spectroscopy taken in the same region \citep{Zaurin08}.  

\begin{figure}[!ht]
	\centering
	\includegraphics[width=14.0cm]{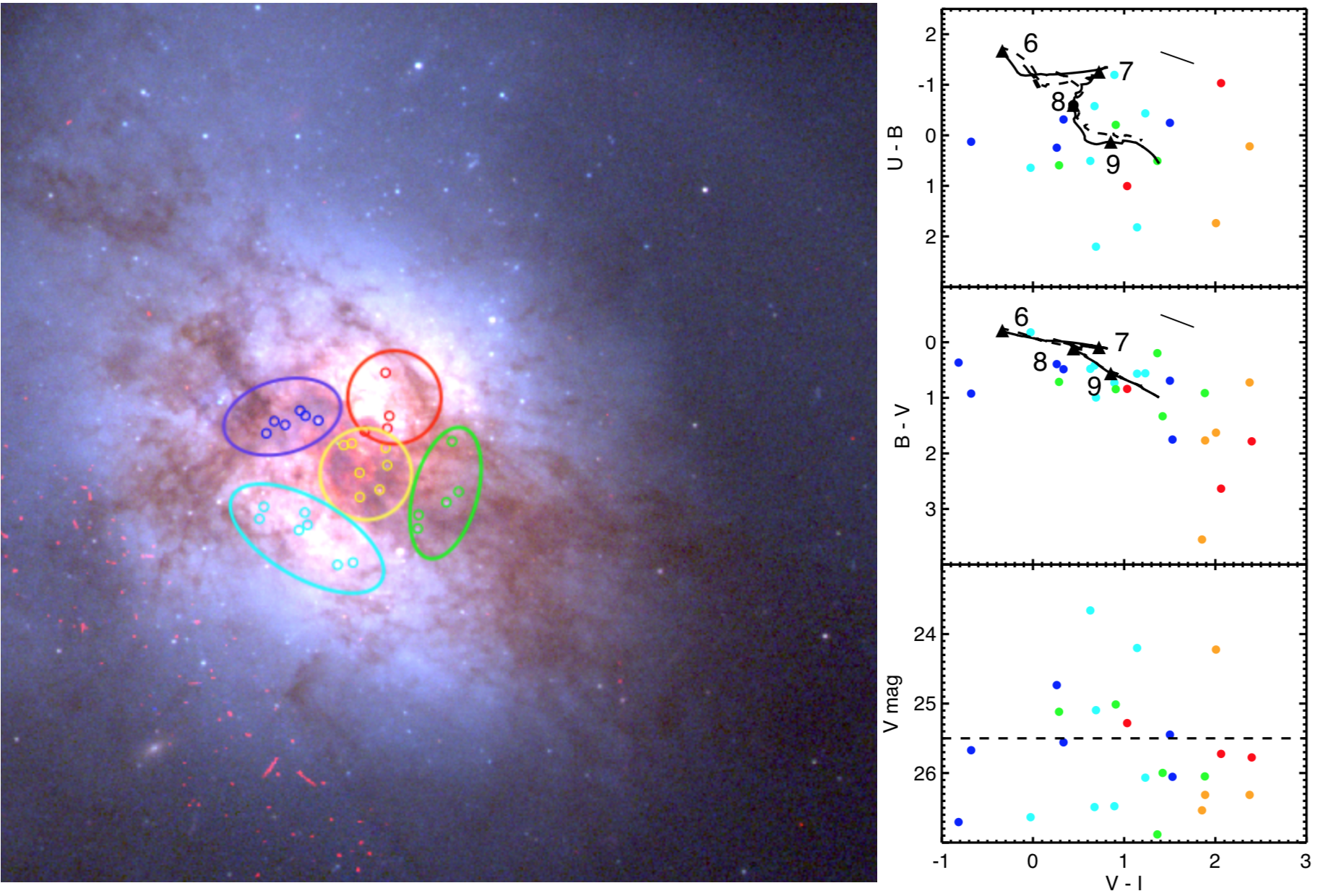}
	\caption{We investigate reddening and extinction in the 5 indicated regions near the center of Arp~220 ($HST$ BVI color image is the backdrop).  The left panel shows the regions and all clusters detected within them, and the right panels show the indicated colors and magnitudes of these clusters.
	See Section~4.1 for details. 
	}\label{fig:colorsreddening}
\end{figure}

We constrain the range of reddening and extinction towards clusters in the central region of Arp~220 by comparing the ranges of measured and expected colors.
In the left panel of Figure~\ref{fig:colorsreddening}, we identify clusters in 5 regions that visually appear to have variable amounts of dust.  
We note that only 9 of the 27 clusters in these 5 regions have a sufficiently bright apparent $V$-band magnitude to appear in Figure~\ref{fig:twocolor}, and that the most reddened clusters (shown in orange and red) comprise a very small fraction (only $\sim1$\%) of the full cluster sample.
For the clusters in these regions, we show the measured U-B vs. V-I (top-right) and B-V vs. V-I (middle-right) colors, along with their V vs. V-I color and magnitude (bottom-right).
The five clusters in the orange circle, which are closest to the center of Arp~220, are clearly the reddest, with all measurements in the top 2 panels falling completely redward of the oldest ages predicted by the BC03 models. If we assume that these clusters have ages of $\sim 100-300$~Myr, they must have E(B-V) values between $\sim 1.5-2.5$~mag, or $A_V\sim 5-8$~mag.
Clusters located in the regions identified by the blue, cyan, and green circles better follow the models than those shown in orange.  If we again assume that they are $\sim100-300$~yr old, they would have E(B-V) values between $\sim 0.0$ and 0.5~mag, or $A_V\sim 0.0-1.5$~mag. The clusters shown in red appear to be intermediate between those in orange and blue/cyan/green, and we estimate E(B-V) value between 0 and 1~mag, or $A_V\sim 0-3$~mag.  These E(B-V) estimates by region allow us to vary the maximum value in our age-dating procedure (described below), based on the location of each cluster.

\begin{figure}[!ht]
	\centering
    \includegraphics[width=12.0cm]{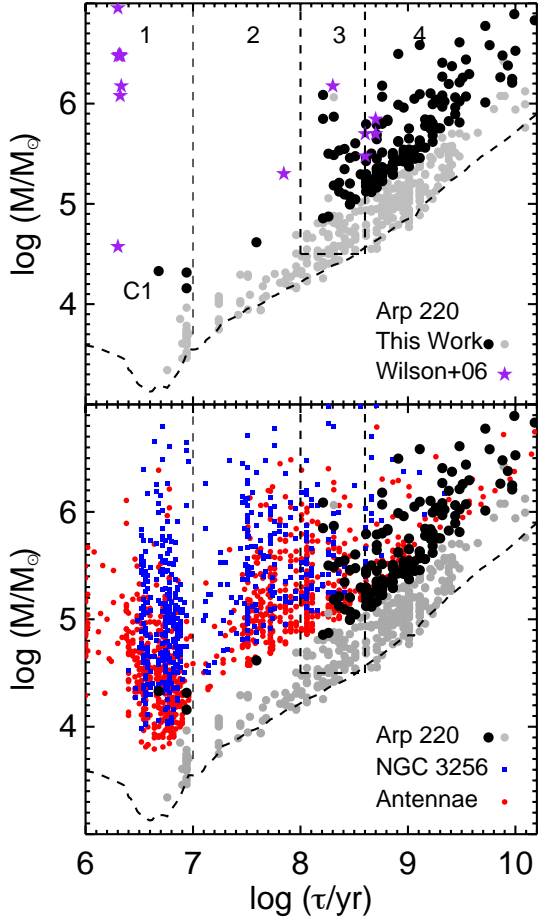}
	\caption{{\bf Top:} Age and mass estimates of the clusters in Arp 220 based brighter (fainter) than $m_V=25.5$ are shown in black (gray). 
	The intermediate age ($100-400$ Myr) clusters studied in Section~5 are indicated, including the adopted completeness limit of log~$(M/M_{\odot})=4.5$.  This limit is defined using the method outlined in Section~\ref{subsec:mf}. The purple stars indicate the age and mass estimates from \citet{Wilson06} for 14 clusters in Arp~220. The clusters which have estimated ages less than 10~Myr from \citet{Wilson06} were incorrectly assigned a very high extinction value, leading to artificially young ages and large masses.  We find these clusters actually have ages of a few$\times100$~Myr and moderate masses. {\bf Bottom:} The age and mass estimates for clusters in Arp~220 are compared with those for clusters in the Antennae (red) and NGC~3256 (blue).}\label{fig-agemass}
\end{figure}

\subsection{Age \& Mass Estimates}

To estimate the masses and ages of the clusters in Arp~220, we follow the general SED-fitting technique described in Chandar et al. (2010), where the measured luminosity of a source in each filter is compared with stellar population model predictions,  but with an updated treatment of the reddening.
We estimate the age ($\tau$) and extinction ($A_V$) for each cluster by performing a least $\chi^2$ fit comparing the observed cluster magnitudes with the predictions from the BC03 population models.  The grid runs over ages from log~($\tau/\mbox{yr})=6.0$ to 10.2, and from $A_V=0.0$ to 0.1~mag for clusters in the outer regions.
For clusters in the inner regions shown in Figure~5, we adopt the following maximum $A_V$ values in the SED fit, based on the discussion above: 1.5~mag for clusters in the blue, cyan, and green regions, 3.0~mag for clusters in the red region, and 7.5~mag for clusters in the orange region. 
\par 
The best-fit values of $\tau$ and $A_V$ minimize the statistic: $\chi^2(\tau, A_V ) = \sum_{\lambda}~W_{\lambda} (m^{\mbox{obs}}_{\lambda}- m^{\mbox{mod}}_{\lambda})^2$ where $m^{\mbox{obs}}$ and $m^{\mbox{mod}}$ are the observed and model magnitudes respectively, and the sum runs over all five broad-band filters NUV, U, B, V, and I. The weight factors in the formula $\chi^2$ are taken to be $W_{\lambda}=[\sigma_{\lambda}^2 +(0.05)^2)]^{-1}$, where $\sigma$ is the photometric uncertainty.  The mass of each cluster is estimated from the observed V band luminosity, corrected for extinction, and the (present- day) age-dependent mass-to-light ratios ($M/L_V$) predicted by the models, assuming a distance modulus $\Delta$(m - M) of 34.72.
\par 
This updated method, which allows for a flexible maximum $A_V$ in the SED fitting based on the amount of reddening in a given region, improves age results in Arp~220. For the dusty central region, we find a number of clusters have best fit ages between 100 and 400~Myr with moderate extinction, consistent with age estimates for a number of clusters in low-extinction areas within the central $12.5\arcsec$. Only the line-emitting clusters identified previously by \citet{Perna20} have best-fit ages younger than 10~Myr. We therefore do not worry about correcting the broad-band photometry for contamination by non-hydrogen emission lines.
Our method of allowing a region-dependent maximum extinction (i.e., a maximum $A_V = 0.1$~mag for clusters outside of $12.5\arcsec$ where there is little evidence of dust, and larger values of $A_V$ for the dusty inner regions) in the SED fitting prevents the common problem of 'catastrophic' age dating for old globular clusters, where they are often best-fit by a too-young age and high reddening \citep{Whitmore20}. 
\par
The largest source of uncertainty in the age estimates comes from uncertainties in the photometry,  particularly in the NUV and $U$ band measurements for fainter clusters.  
These uncertainties tend to be largest for the clusters which fall furthest from the models in Figure~\ref{fig:twocolor}.  
This method typically produce uncertainties at the level of log~$(\tau/\mbox{yr}) \approx0.3$ ($\approx$factor of 2); see e.g., \citet{Chandar10b}.
The uncertainties in the mass estimates are tied to those in the ages, and also are on the order of log~$(M/M_{\odot}) \approx0.3$. There can also be systematic errors in the estimated masses that depend on  uncertainties in the assumed stellar initial mass function (IMF) and distance to Arp~220. An error in the estimated distance affects all masses in the same way, and does not impact the shape of the cluster mass function. Several studies have shown that accounting for stochastic variations in the stellar IMF can introduce systematic uncertainties in the derived ages and masses of individual clusters \citep[e.g.,][]{Krumholz15}. However, as \citet{Fouesneau10} demonstrate, these differences are small for clusters with ages $\tau \gea 100$ Myr (and masses $M \gea 10^{4} M_{\odot}$) as we are dealing with in Arp~220. We therefore do not expect stochasticity in the stellar IMF to significantly affect our age and mass results.

\par 
In Figure~\ref{fig-agemass}, we plot our age-mass results for the clusters in Arp~220.
The approximate completeness limit is shown by the solid curve along the bottom of the data points (and defined in Section~\ref{subsec:mf}). 
We find there are almost no detected clusters younger than log~$(\tau/\mbox{yr}) \lea 8.0$,
and none with believable age estimates between log~$\tau=7-8$ and brighter than $m_V\lea 25.5$~mag. 
The single cluster that has a V-band magnitude brighter than 25.5~mag and log~$\tau \approx7.5$ has no B or I band photometry, leading to significantly larger uncertainty in the estimated age than is typical.  The handful of clusters with best-fit ages younger than 10~Myr have very low estimated masses.  C1 is the most massive in this interval, with an estimated mass of $\sim2\times10^{4}~M_{\odot}$ at an age of $\sim3$~Myr. 
\par 
One critical question is: are we missing clusters younger than 100 Myr with masses $\gea 10^5~M_{\odot}$ due to dust obscuration?
To address this question, we carefully examined our H$\alpha$ and P$\beta$ line maps for any additional line-emitting clusters besides C1-C4, but did not find any.
While we cannot firmly rule it out, the multi-wavelength observations presented here along with previous deep infrared and radio observations suggest that massive ($\gea 10^5~M_{\odot}$) and  young ($\lea 6$~Myr) line-emitting clusters are not 'hiding' in the central $12.5\arcsec$ of Arp~220 (although there may be lower mass ones).  
There is also circumstantial evidence against massive clusters with ages between $\sim7$ and 100~Myr having formed in Arp~220.  The region outside of $12.5\arcsec$ has little dust, and there are no clusters younger than 100~Myr in this area, the majority of Arp~220.  Within $12.5\arcsec$, approximately half of the area has no visible dust lanes and the clusters in these regions have colors which suggest ages $\gea 100$~Myr, with none between $7-100$~Myr.  If there are reddened, $7-100$~Myr clusters in Arp~220, they must be located {\em only} in the dusty regions within the inner $12.5\arcsec$. 
\par 
The general lack of clusters younger than $\tau \lea 100$~Myr in Arp~220 is consistent with expectations of a post-starburst system. In the mass-age diagram, clusters appear abruptly around this age, then continue more-or-less continuously to older ages. While the age-metallicity degeneracy makes it challenging to precisely age date clusters older than log~$(\tau/\mbox{yr}) \approx 8.6$ ($\approx400$~Myr), cluster formation in Arp~220 appears to have been fairly continuously between 100~Myr and the past several Gyr.

\par
In the bottom panel of Figure~\ref{fig-agemass}, we compare the overall age-mass demographics of clusters in Arp~220 with those in two on-going merging systems, the Antennae and NGC~3256.  The cluster system of Arp~220 is quite different from the others, which have formed massive ($\gea 10^5~M_{\odot}$) clusters more-or-less continuously over the past several hundred Myr, including in just the last few million years.  Spiral galaxies like M51 and M83, dwarf starbursts like NGC~4449 and NGC~4214, as well as irregular galaxies like the LMC and SMC also have continuous cluster age-mass diagrams \citep{Chandar17}.
The optically visible clusters in Arp~220, including those in the central kpc, are almost all older.  This means that if young clusters have formed in the highly obscured, $\approx 150$~pc nuclear disks, none have diffused out.

\subsection{Comparison with the Wilson$+$06 Cluster Catalog}
\begin{figure}[!ht]
	\centering
	\includegraphics[width=8cm]{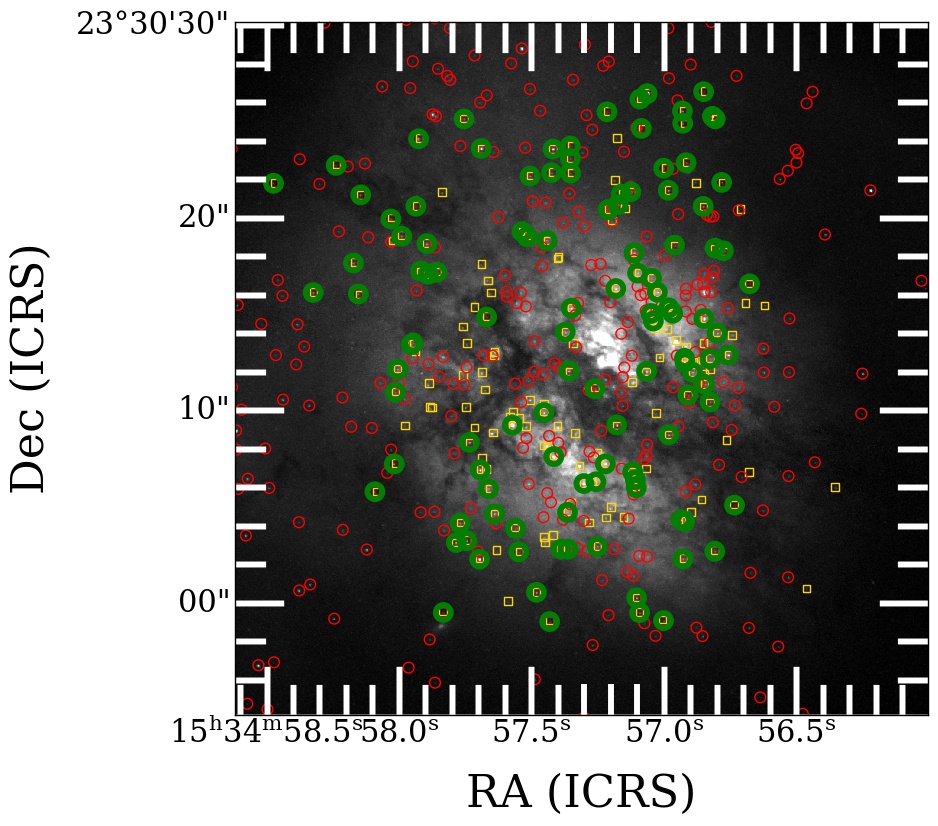}
	\includegraphics[width=8cm]{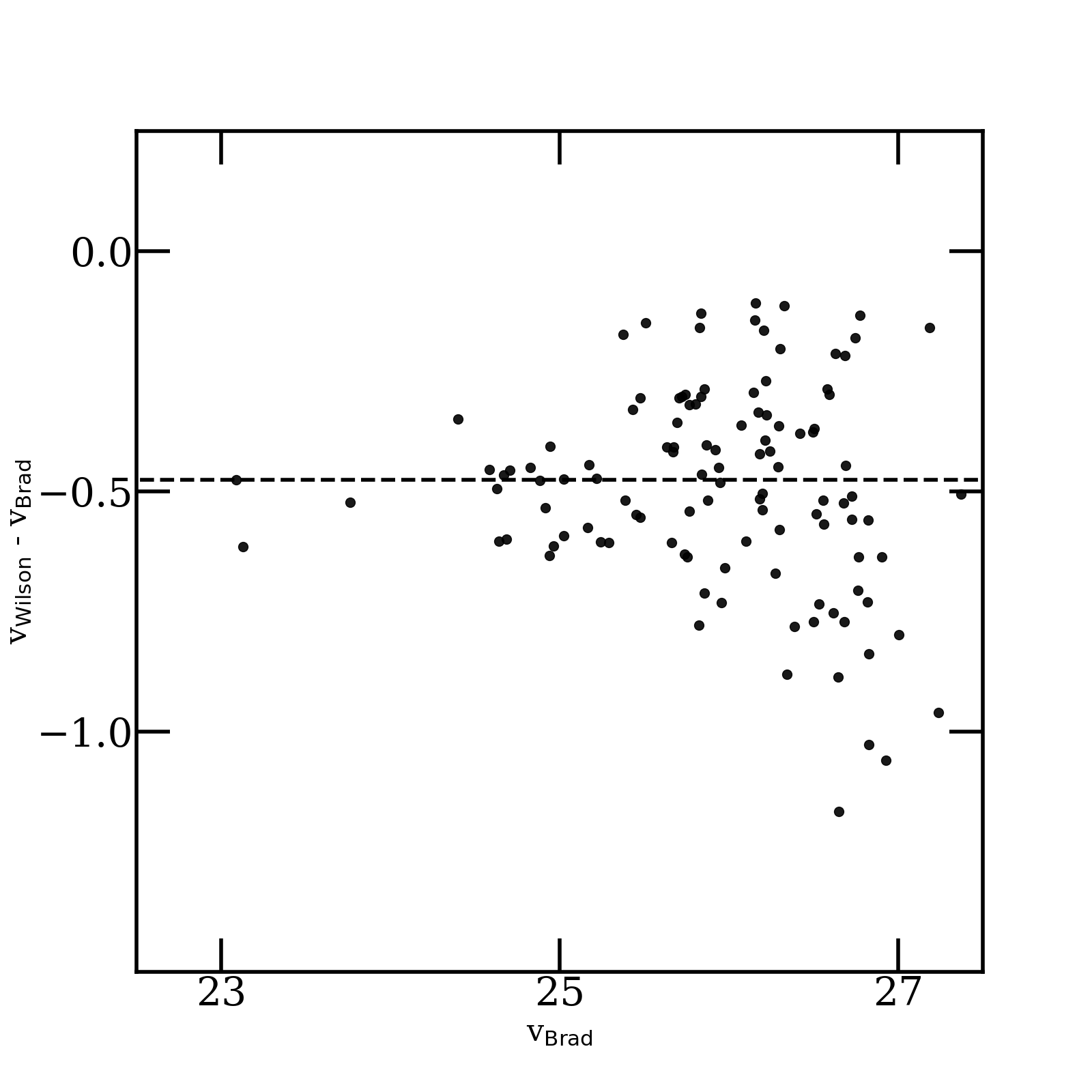}
	\caption{{\bf Left:} Clusters detected in this work (red) and those detected by \citet{Wilson06} (yellow squares) are shown. Objects in common are outlined in green. {\bf Right:} A comparison of the V-band magnitude from our Arp~220 cluster catalog compared with that listed in \citet{Wilson06}. We find a median difference of $-0.476$~mag, such that our magnitudes are fainter.}\label{fig-vcomp}
\end{figure}
\par
\citet{Wilson06} published one of the few studies of the cluster population in Arp 220, based on $UBVI$ images of the central $26\times29\arcsec$ taken with the ACS/HRC camera on $HST$.
They detected 206 candidate clusters (down to $m_I\approx$ 25~mag), shown as the yellow squares in Figure~\ref{fig-vcomp}a.  The 
vast majority ($\sim85$\%) of these sources are also in our catalog. \citet{Wilson06} were able to
estimate the ages and masses for 14 clusters by comparing their measured colors with predictions from the \citet{Bruzual03} models. Seven of these clusters have $UBVI$ photometry, while the other seven only have $BVI$.  
\par
We match the locations of clusters from our catalog with those from \citet{Wilson06}, and show the overlap between the two catalogs in the left panel of Figure~\ref{fig-vcomp}. We detect the majority of the clusters in their catalog (sources in common are circled in green), and detect many fainter clusters as well (see the many sources circled in red).
We compare the total $V$ band magnitude of the matched clusters in the right panel of Figure~\ref{fig-vcomp}.  There is a median offset of $-0.48$~mag between the two works, with the \citet{Wilson06} luminosities being brighter; we are uncertain of the origin of this offset, but note that it is challenging to determine aperture corrections in the crowded, highbackground region used by \citet{Wilson06}.
\par
Similar to our findings here, 
\citet{Wilson06} find only one cluster (C1) with a very blue $U-B$ color.  For the other 6 clusters with $U$ band photometry in their catalog, they estimate ages of a few $\times100$~Myr.  However, for the 7 clusters which they do not detect in the $U$ band, they {\em assume} ages of $1-3$~Myr, which results in fairly large reddening values of $E(B-V)\approx 1-2$~mag and artificially boosts the estimated masses above a million solar masses.

\section{Shape of the Cluster Mass Function} \label{subsec:mf}

\subsection{Results for Arp~220}

\begin{figure}[!ht]
	\centering
	\includegraphics[width=7.5cm]{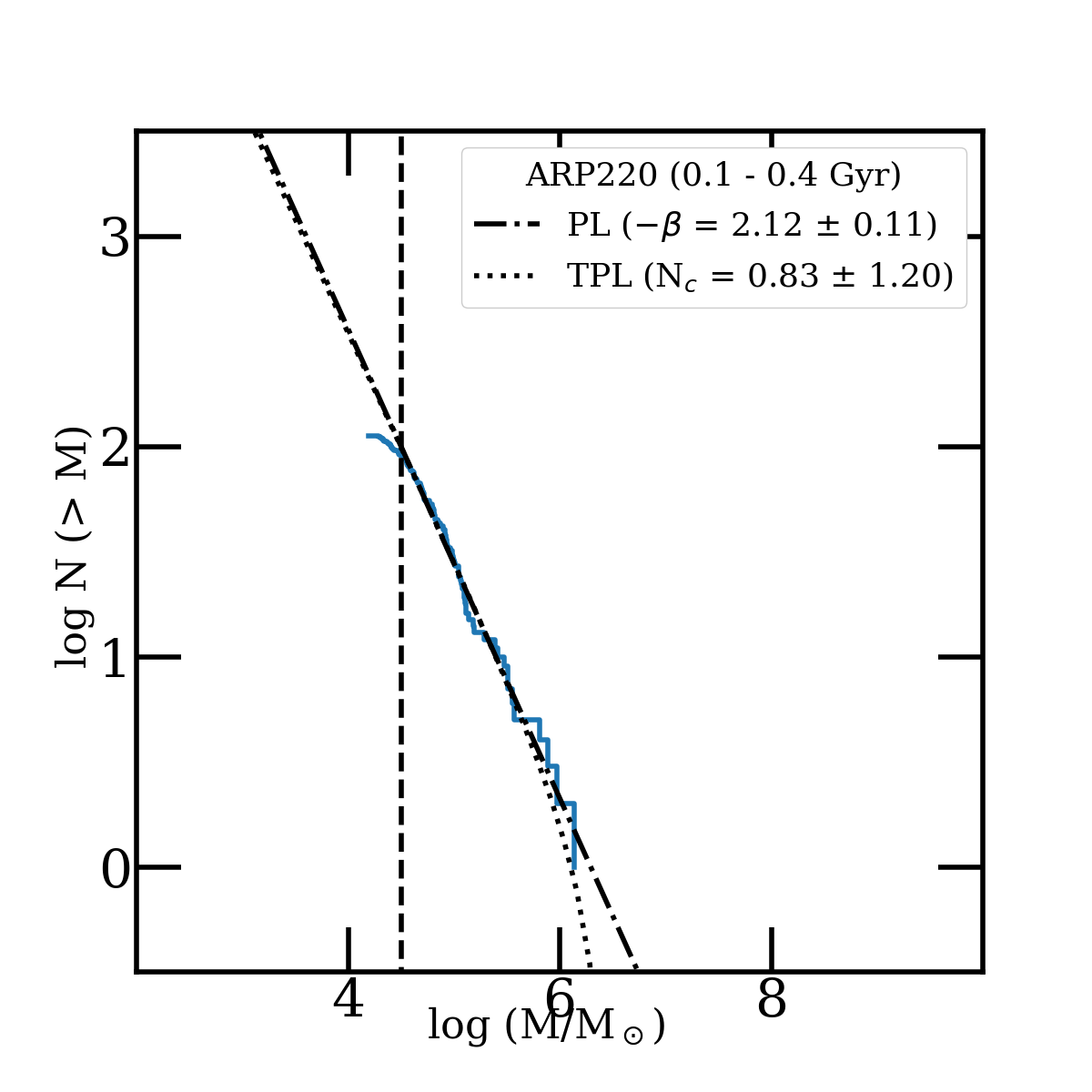}
	\includegraphics[width=7.5cm]{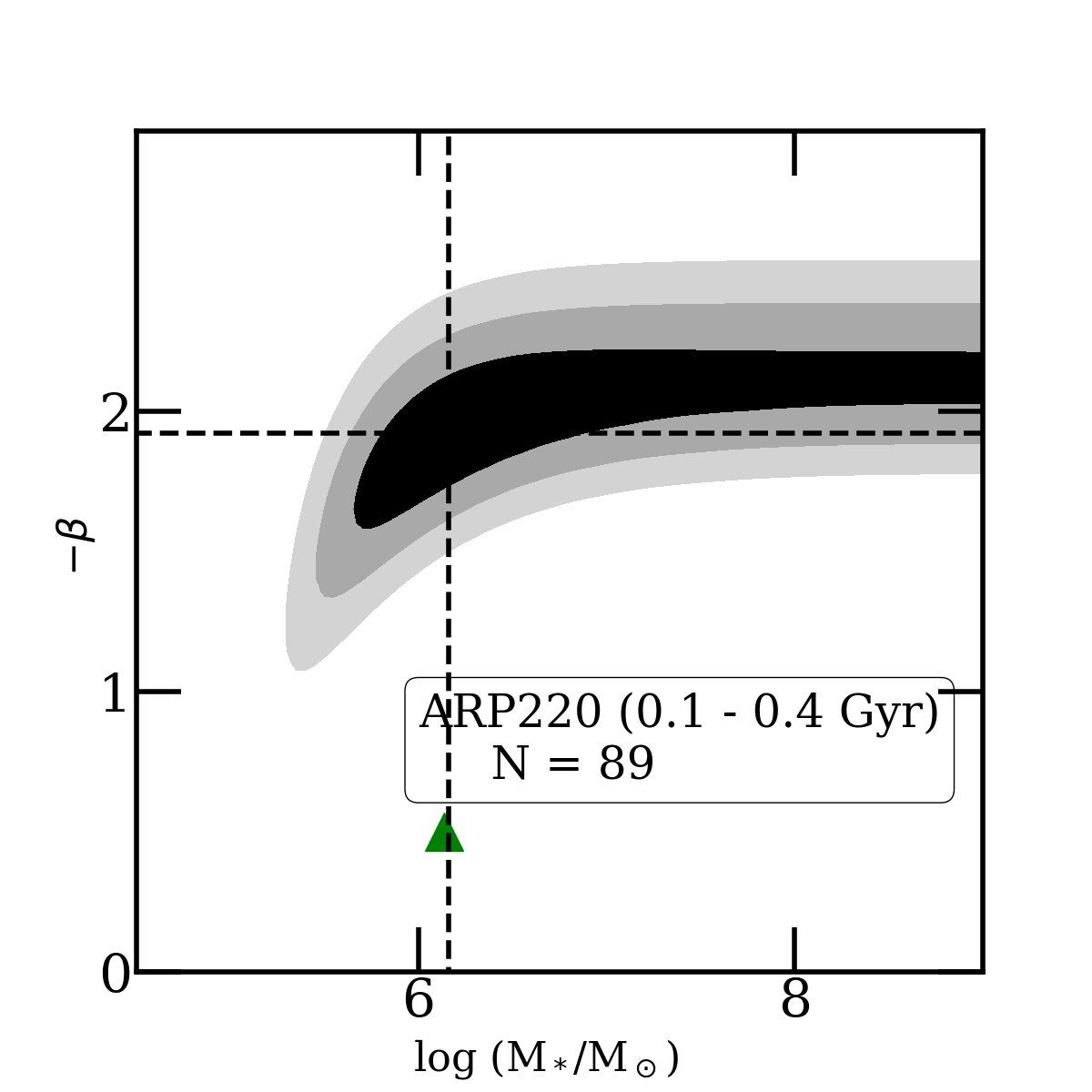}	
	\caption{{\bf Left:} Cumulative cluster mass function for 100 - 400 Myr clusters in Arp~220. The completeness limit (vertical dashed line) is set where the mass function deviates from a power law by a factor of 2. {\bf Right:} Maximum likelihood confidence contours ($1\sigma$, $2\sigma$, and $3\sigma$) of the Schechter function fit to the mass function of the clusters. See text for details. }\label{fig-cdf}
\end{figure}

\par
The shape of the cluster mass function, particularly whether or not there is a physical cutoff or truncation at the upper end, provides important information on the formation of the clusters. In this section, we use clusters formed over the 100-400~Myr age interval to assess whether there is a physical (or just a statistical) cutoff in the cluster mass function in Arp~220.  This age range is ideal in general because clusters at these ages can be reliably detected, classified, and age-dated, and the interval covers a fairly long elapsed time which increases the available statistics. After $\sim400$~Myr, clusters are affected by the age-metallicity degeneracy.  In Arp~220, almost no clusters have formed in the last 100~Myr, so it is not possible to use younger age intervals. 

\par 
We present the cumulative mass function for 100-400~Myr clusters detected throughout Arp~220 in Figure~\ref{fig-cdf}-left.  
This figure shows that the mass distribution follows a power law at the upper end, but eventually flattens towards lower masses. We assume that this flattening is due to incompleteness, rather than to a physical effect, just as we have done previously for cluster catalogs in other systems \citep[e.g.,][]{Chandar17, Mok19}. 
We set the completeness limit $M_{\rm lim}$ at $\log M /~M_\odot = 4.5$, the mass where the distribution begins to flatten noticeably, i.e. where the mass function begins to fall significantly below the extrapolated power-law (by a factor of $\sim2$), represented by the dashed line in the Figure.
\par 
We use the maximum-likelihood method described in \citet{Mok19} to determine the best-fit value and confidence intervals for $\beta$ and $M_*$ when fitting a Schechter function to the cluster masses above the completeness limit, $\psi(M) \propto M^{\beta}~\mbox{exp}(-M/M_*)$. This method does not use binned data, which can hide weak features at the ends of the distribution,  or cumulative distributions, where the data points are not independent of one another. We compute the likelihood $L(\beta,M_*)=\Pi_i P_i$ as a function of $\beta$ and $M_*$, where the probability $P_i$ for each cluster is given by:
\begin{equation}
P_i = \frac{\psi(M_i)}{\int_{M_{\rm min}}^{\infty} \psi(M)dM} 
\end{equation}
and the product is over all clusters above the completeness limit (see e.g. Chapter~15.2 of \citet{Mo10}).  We set the upper integration limit in equation (1) to be 100 times the mass of the most massive cluster, which is sufficient for convergence.  Next, we find the maximum-likelihood $L_{\rm max}$ using the \citet{Nelder65} method, and use the standard formula:
\begin{equation}
\ln L(\beta, M_*) = \ln L_{\rm max} - \frac{1}{2} \chi_p^2(k)
\end{equation}
where $\chi_p^2(k)$ is the chi-squared distribution with $k$ degrees of freedom at $p$ confidence level to determine the 1-, 2-, and 3-$\sigma$ confidence contours.
\par 
Figure~\ref{fig-cdf}-right shows the best-fit values of $\beta$ and $M_*$ (dashed lines) for the $100-400$~Myr clusters in Arp~220. The shaded regions show the 1, 2, and 3$\sigma$ contours resulting from our maximum-likelihood fit. The contours show a short diagonal portion,  which indicates the trade-off between a steeper value of $\beta$ and a higher cutoff mass $M_*$, and then a relatively flat portion up to the highest mass tested.  The green triangle shows that the most massive cluster in the sample has a mass very similar to the best fit $M_*$. The fits return a lower limit for $M_*$ of $\approx 5\times10^5~M_{\odot}$ at the 95\% confidence level.
If there is a significant detection of a physical cutoff, the $3\sigma$ contours will close within the plotted $\beta-M_*$ parameter space.  For Arp~220, even the $1\sigma$ contour remains open all the way up to the maximum tested mass, indicating that the detected $M_*$ is not significant even at the $1\sigma$ level.
We find a best fit index of $\beta=-2.12\pm0.11$ when we fit a pure-power law, by allowing $M_* -> \infty$, and $\beta=-1.93 \pm 0.14$ when marginalized over all $M_*$.
{\em These results mean that the value of $M_*$ returned by the maximum likelihood fit is indeterminant and represents a lower limit to the actual value; the upper portion of the mass function is consistent with a pure power law with $\beta=-2.12\pm0.11$.} 

\subsection{Comparison with Predictions and Previous Work}

The presence or absence of an upper cutoff in the cluster mass function figures prominently in the input and output of different cosmological simulations \citep[e.g.,][] {Li17,Li18}, and in understanding star and cluster formation from progenitor molecular clouds \citep{Grudic21}. Disruption is not expected to affect the high mass end of the mass function over the first 400~Myr, so we attribute this shape to {\em formation} rather than to disruption.

\par 
The observational literature characterizing physical cutoffs has been decidedly mixed. The most compelling evidence for a physical cutoff is found in the nearby spiral galaxy M31 \citep{Johnson17}, from the Panchromatic Hubble Andromeda Treasury (PHAT) program that surveyed $\sim1/4$ of the galaxy disk (but did not include the regions of strongest star formation). A recent extension of the PHAT survey to M33 has also provided evidence of a physical cutoff at $\sim 10^{4} M_{\odot}$ \citep{Wainer22}.
There are also contradictory claims for cutoffs in the young cluster populations in M51 \citep{Messa18a}, and M83 \citep{Bastian12, Adamo15}, some of them based on the same data. Other groups have found no preference for a physical cutoff in NGC~4449 \citep{Whitmore20}, in a sample of 7 out of 8 nearby galaxies \citep{Mok19}, in a composite of 17 dwarf galaxies studied as part of the LEGUS survey \citep{Cook19}, and for 2 out of 6 LIRGs in the Hi-PEEC Survey \citep{Adamo20}. Our results for Arp~220 add to the growing list of galaxies with little evidence for a physical cutoff. 

A cosmological simulation by \citet{Li17} developed a new algorithm for modelling the formation and growth of clusters in Milky-Way size galaxies at high redshift.  The clusters grow over several million years, until halted by energy and momentum feedback.  These simulations result in initial cluster mass functions that are better described by a Schechter function than a single power law, with an upper cutoff that scales with the star formation rate.   For their fiducial run, \citet{Li17} find a strong correlation between the predicted upper cutoff ($M_*$), the most massive cluster ($M_{\rm max}$), and SFR:
\begin{equation}
M_* \approx  1.4\times 10^4 \left( \frac{SFR}{1M_{\odot}~\mbox{yr}^{-1}} \right)^{1.6},
\end{equation}

\begin{equation}
M_{\rm max}\approx 8.8\times 10^4 \left( \frac{SFR}{1M_{\odot}~\mbox{yr}^{-1}} \right)^{1.4}.
\end{equation}

\noindent Different assumptions for the simulation parameters related to cluster formation and feedback somewhat affect their quantitative results for $M_*$ and $_{\rm max}$, but not the fundamental finding that the mass function is better represented by a Shechter function than by a power-law.
A related trend in the \cite{Li17} simulation is that a higher fraction of stars are found in bound clusters when there is a higher $\Sigma_{SFR}$.
\par 
Here we compare predictions from the fiducial
\citet{Li17} model with our results for Arp~220. At a SFR of $5~M_{\odot}~\mbox{yr}^{-1}$ ($10~M_{\odot}~\mbox{yr}^{-1}$) which covers the range we find for Arp~220 in Section~6.4 for the 100-400~Myr age interval, the model predicts a cutoff (Schechter) mass of $M_*\approx1.8\times 10^5$ 
($M_*\approx5.6\times 10^5$)
and a maximum mass of
$M_{\rm max}\approx 8.4\times 10^5$ 
($M_{\rm max}\approx 2.2\times 10^6$).
For these SFRs, $M_{\rm max} \approx 4\times M_*$.  
In Section~4, we found that the 100-400~Myr clusters in Arp~220 do not show statistically significant evidence for a cutoff mass, with a formal best fit value of $M_*\approx 1-2\times10^6~M_{\odot}$, almost identical to the value of $M_{\rm max}$ for the sample.

\citet{Li18} presented improvements to the formation and feedback prescriptions used in their models. They find a relation between the maximum cluster mass and the SFR density of $M_{\rm max} \sim \Sigma_{SFR}^{2/3}$, where the range of normalizations for $M_{\rm max}$ is shown in their Figure~13.  We estimate a $\Sigma_{\rm SFR}\approx0.01$ in the interval 100-400~Myr ago, based on a SFR$\approx10$ and a box with the area enclosed by $40\arcsec$ (see Figure~1\ref{fig:arp220}). Their Figure~13 implies $M_{\rm max} \sim \mbox{several}\times10^4~M_{\odot}$ for this $\Sigma_{\rm SFR}$, significantly lower than our results for Arp~220, likely because their runs have a strong upper cutoff at the high mass end. Other theoretical work, such as simulations of star and cluster formation within turbulent, star-forming giant molecular clouds (GMCs) which include stellar feedback, also find a Schechter-like shape for the initial cluster mass function \citep{Grudic21}.  It will be important for future simulations to  provide ranges of model parameters that can accommodate a wide range of initial cluster mass functions, including power-law ones.

\section{Cluster-Based Estimates of the Star Formation Rate in Arp~220}

In this Section we estimate or constrain the rate of star formation in different intervals of age based on measured properties of the clusters.

\subsection{Methods}

Building on earlier work \citep[e.g.][]{Bastian08,Chandar21}, here we develop two cluster-based calibrations to estimate the SFR in the parent galaxy. The mass function of young cluster populations in nearby galaxies ($\tau \lea 0.5$~Gyr) have a near-universal shape, which is well-described by a power law, $dN/dM \propto M^{\beta}$, with $\beta=-2.0\pm0.2$ \citep[e.g.][]{Zhang99,Fall12}, and scales with the total SFR (Chandar et al. 2017; see below).  This near-universality in the relationship between the cluster mass function and SFR also means that the most massive cluster (or the 3rd, 5th, etc) can serve as a proxy for the normalization of the mass function \citep{Chandar21}, as long as the upper end of the cluster mass function is shaped by statistics rather than a physical truncation.

\begin{figure}[!ht]
	\centering
	\includegraphics[width=8.5cm]{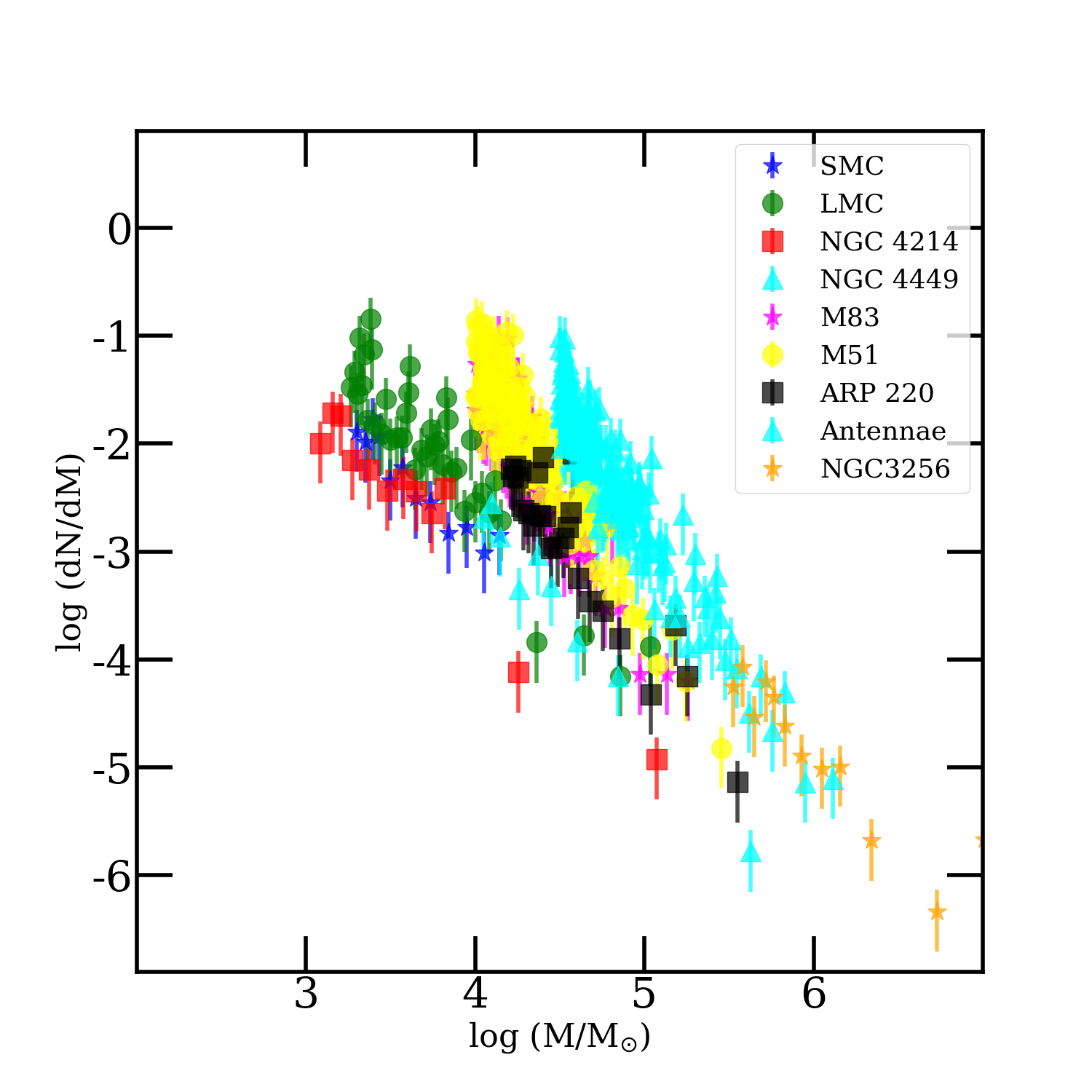}
	\includegraphics[width=8.5cm]{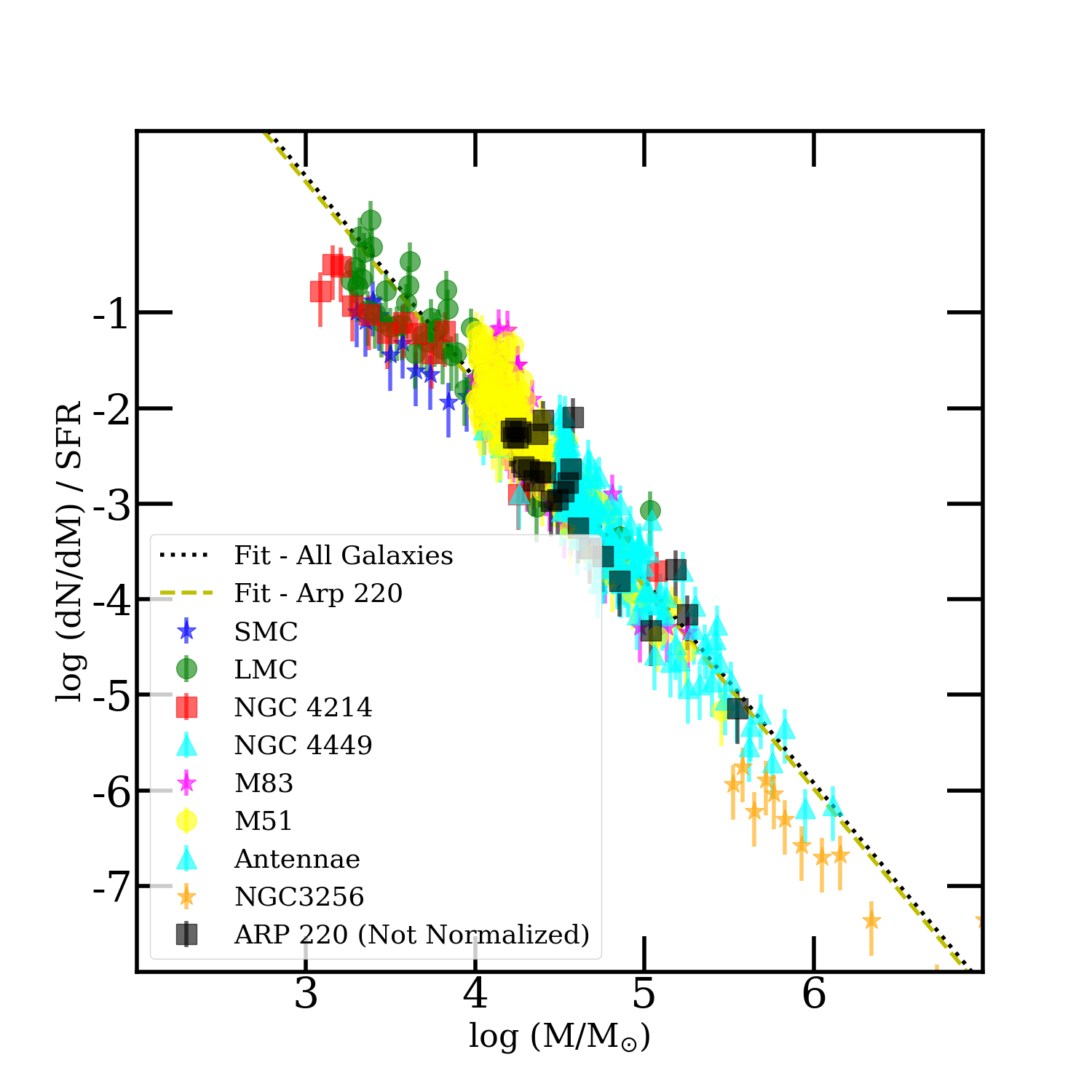}
	\caption{Demonstration of Method~1 for estimating the SFR.  {\bf Left:} The cluster mass function (CMF) for $100-400$~Myr clusters in the eight galaxies analyzed in \citet{Chandar17} and for Arp 220 (black squares) are shown using an equal number of objects per bin. The legend shows the symbol used to represent each galaxy. {\bf Right:} The right panel shows the CMF divided by the SFR, along with the best fit CMF/SFR for clusters in Arp~220.  The power-law fits to the calibration galaxies and to the CMF$/$SFR distribution for Arp~220 are very similar, and yield a best fit SFR of $2.4~M_{\odot}~\mbox{yr}^{-1}$. 
 See text for details.} \label{fig-Arp220cmfsfr}
\end{figure}

Previously, we showed that the observed CMF in 8 galaxies is proportional to the SFR of the host galaxy.
This is demonstrated in Figure~\ref{fig-Arp220cmfsfr}.  In the left panel, we show the observed mass function for clusters in the SMC and LMC (irregulars), NGC~4214 and NGC~4449 (dwarf starbursts), M83 and M51 (spirals), Antennae and NGC~3256 (mergers), in the age interval between 100 and 400~Myr.  
The global SFRs in these  galaxies (compiled in Table~1 in \citet{Chandar17}) span a factor of nearly 1000, which is reflected in their very different vertical scales.
The right panel shows that when divided by the SFR, the mass functions for these very different galaxies collapse to form essentially a single relation.  A 'universal' CMF$/$SFR sequence means that we can estimate the star formation rate of the host galaxy from the observed CMF.
\par 
Based on this result, we develop two cluster-based methods to estimate the star formation rate in nearby galaxies. Our methods have the advantage that the SFR can be determined in different intervals of age, which can be adapted to the system being studied.  For Arp~220, we will focus on the $1-10$~Myr, $10-100$~Myr, 100-400~Myr, and 400~Myr$-$3~Gyr age intervals. 
This ability to select specific age intervals presents a distinct advantage over global star formation rate tracers, such as extinction-corrected far ultraviolet or H$\alpha$ luminosities.

\begin{figure}[!ht]
	\centering
\includegraphics[width=12.0cm]{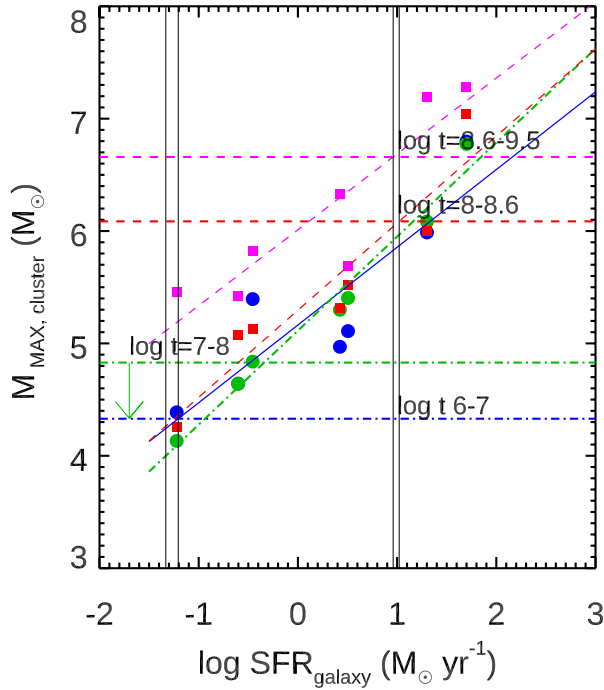}
	\caption{Demonstration of Method~2 for estimating the SFR. This figure shows the calibration between SFR and the most massive cluster ($M_{\rm max}$), based on the 8 galaxies (LMC, SMC, NGC~4214, NGC~4449, M83, M51, Antennae, and NGC 3256) studied in \citet{Chandar17}. The calibration is done separately for each of four age intervals that we study in Arp~220: (1) log $(\tau/\mbox{yr}) =6-7$ (blue circles), (2) log $(\tau/\mbox{yr}) =7-8$ (green circles),  (3) log $(\tau/\mbox{yr}) =8-8.6$  (red squares), and (4) log $(\tau/\mbox{yr}) =8.6 - 9.5$ (pink squares). The color-coded diagonal lines show the best fit relation for the calibration galaxies, and the most massive cluster found in Arp~220 in each age interval is indicated by the horizontal line with the appropriate color. See text for details.}\label{fig-MmaxSFR}
\end{figure}

\begin{itemize}

\item {\bf Method~1:} This is the more robust method, but requires a sufficient number of clusters to determine the mass function.
We fit a power law to the combined CMF$/$SFR distribution of the eight calibration galaxies listed above, using a least-squares fitting routine. Then, a power law with the same slope is fitted to the cluster mass function of the galaxy for which we want to estimate the SFR.  We highlight that the exact same age interval must be used for the calibration and SFR determination. The best fit normalization required to match the mean CMF$/$SFR statistic from the eight galaxies gives the estimated SFR.  The method is demonstrated in Figure~\ref{fig-Arp220cmfsfr}.

As a test of our method, we applied the procedure to the observed mass function of 100-400~Myr clusters in the Antennae (after first taking this galaxy out of the calibration).  Our procedure returns an estimated SFR of  $12.5~M_{\odot}~\mbox{yr}^{-1}$, quite similar to the $11\pm3~M_{\odot}~\mbox{yr}^{-1}$ estimated from global measurements of the far ultraviolet and infrared luminosity for the Antennae (see Table~2 in \cite{Chandar17}).  We will apply Method~1 to estimate the SFR in the $100-400$~Myr age interval in Arp~220.

    \item  {\bf Method~2:}
A direct consequence of the proportionality between the CMF and SFR  is that the total number of clusters, and therefore the mass of the most massive cluster, also scales directly with the SFR of the host galaxy \citep[e.g.][]{Whitmore14,Chandar15}.
When the cluster sample is insufficient to determine the mass function, the method developed in \citet{Chandar21} and applied to a post-starburst galaxy can be used.  This method uses the best fit from the calibration between the first (3rd, 5th) most massive cluster in a given age interval and the SFR, and is demonstrated in Figure~\ref{fig-MmaxSFR}.   This method has significantly larger uncertainties than Method~1, but the advantage that it can be applied when only a handful of clusters are detected.  For the Antennae, we find estimates between $9-14~M_{\odot}~\mbox{yr}^{-1}$ for the SFR when we apply Method~2 using the 1st, 3rd, and 5th most massive cluster.  We will use this method to constrain the SFR in the following age intervals in Arp~220: $1-10$~Myr, $10-100$~Myr, $0.4-2$~Gyr.  

\end{itemize}

Here, we perform a sanity check that it is reasonable to apply Method 1 to Arp 220. First, we note that over the $100-400$~Myr age interval, clusters in Arp~220 appear to have formed fairly continuously, just as in the galaxies used for the calibration between SFR and CMF.  We have also measured the fraction of light found in intermediate age clusters in the region between $12.5-45\arcsec$, where there is little dust.  We find that $\sim 2-3$\% of the light in this region is found in clusters, quite similar to the result found by \citet{Whitmore20} for regions in NGC~4449 that are dominated by a similar age population.

\par 
\subsection{Star Formation Rate Estimates Outside of Nuclear Disks}
In this section, we use the two methods described above to estimate the star formation rate using observed cluster masses in different age intervals.

\begin{itemize}

\item {\bf  Past $10$~Myr:} Here we estimate the current rate of star formation outside of the nuclear disks using Method~2. Our optical sample only contains two clusters (C1 and C2) that are younger than 10 Myr in the central $12.5\arcsec$.  C1 is the brightest and has the highest estimated mass, which is $M\approx 1-2\times10^4~M_{\odot}$. This gives an estimate of the SFR of $0.06~M_{\odot}~\mbox{yr}^{-1}$, similar to the current star formation rate in the SMC.

\item {\bf  $10-100$~Myr:} There are essentially {\em no} clusters in our catalog with ages between 10 and 100~Myr, as discussed in Section~4.2.\footnote{ 
Two clusters in the outer portion of Arp~220 which have somewhat blue $U-B$ colors are quite faint in the $U$ band and have large photometric uncertainties.  These are consistent with ages $\gea 100$~Myr in the $B-V$ vs. $V-I$ color-color diagram.  No clusters in the inner $12.5\arcsec$ of Arp~220 have colors which suggest ages between 10 and 100~Myr.}
Previous spectroscopic studies of Arp~220 did not discriminate between stellar population ages between 1 and 10~Myr and those between 10 and 100~Myr (e.g., Rodriguez Zaurin et al. 2008, 2009), so do not provide any additional constraints on the rate of star formation during the 10-100~Myr interval specifically.  

We use the lack of clusters in this age interval to set an upper limit on the rate of star formation.  The luminosity limit of clusters in our catalog (dashed curved line along the bottom of the data points in Figure~\ref{fig-agemass}) indicates that we should be able to detect clusters down to $\sim10^4~M_{\odot}$ over the $10-100$~Myr age interval. Using Method~2, this provides an upper limit for the SFR of $0.05~M_{\odot}~\mbox{yr}^{-1}$.

\item {\bf  $100-400$~Myr:} Here, we use Method~1 to estimate the SFR of Arp~220 over the 100 to 400~Myr age interval. The cluster mass function in equal-number bins is shown as the black squares in Figure~\ref{fig-cdf}-left.  
The observed CMF falls closest to those for the spirals M83 and M51.
Using Method~1, we find a best fit normalization of $2.4~M_\odot$ yr$^{-1}$ to the CMF$/$SFR distribution from the other galaxies, with no correction for   incompleteness in the dusty inner $12.5\arcsec$. Based on the area covered by the dusty regions and number of clusters detected in the inner region, we estimate a 20\% correction, giving an estimated SFR of $\approx3~M_\odot$ yr$^{-1}$.
If we use Method~2 instead, we find estimates in the range 4 to 9~$M_{\odot}~\mbox{yr}^{-1}$ from the 1st, 3rd, and 5th brightest cluster.
Based on these results, we estimate that the SFR between 100 and 400~Myr ago was between  and $\sim3-9~M_{\odot}~\mbox{yr}^{-1}$.

\item {\bf $>400$~Myr:} Arp~220 has formed many clusters which have sufficiently red colors (in areas of low extinction) that they are almost certainly older than 400~Myr. While there is no known calibration between star formation rate and cluster masses at these ages,  we can place some loose (but uncertain) constraints on the star formation rate by comparing the properties of the clusters that fall redward of the 400~Myr model age to cluster populations found in typical irregular, spiral, and other merging galaxies.  By applying the procedure used in Method~2 to this age range, we estimate a SFR on the order of $\approx10~M_{\odot}~\mbox{yr}^{-1}$. 

\end{itemize}

\section{Discussion}\label{sec:discussion}

\subsection{Gas, ISM, and the Lack of Current Star Formation Outside of the Nuclear Disks}
\par 
Molecular gas emission in Arp~220 originates from the two, $\sim100$~pc-scale nuclear disks and also from a larger, kpc-scale disk \citep[e.g.][]{Scoville97,Sakamoto99,Sakamoto08,Rangwala15,Wheeler20}. Approximately half of the molecular gas is found in the compact disks and the other half in the extended components, while $\sim10$\% of this gas is warm as traced by CO 3-2 \citep{Rangwala11}. Approximately half of the total infrared \citep[e.g.][]{Dwek20} and P$\beta$ emission (this work) also originates from the extended disk. 

A number of studies have unsuccessfully searched for signatures of on-going star formation in the extended kpc-disk region. If P$\beta$ emission traces star formation, we would expect to see a comparable number ($\sim100$) of radio point sources {\em outside} of the nuclear disks as found within them.

Yet previous searches for deeply embedded sources in the radio and infrared have not detected any point sources in this extended region \citep{Varenius19}, including a dedicated infrared supernova search (J. Kezwer thesis, https:$//$dspace.library.uvic.ca$/$handle$/$1828$/$4992). 
These results at longer wavelengths are consistent with the lack of detections of young clusters in the optical/IR HST observations.

Rotational transitions are an excellent tracer of the physical conditions of the gas over a range of temperatures (10–1000 K) and densities ($10^3-10^8~{\rm cm}^{-3}$), and can be used as diagnostics to differentiate between different energy sources responsible for exciting the gas.
ALMA and Herschel SPIRE-FTS observations of the warm molecular gas in Arp~220, which dominates the CO emission, suggests it is impacted by mechanical energy rather than by photoionization \citep{Rangwala15}.
 This is similar to the situation in some post-starburst galaxies, which can also have significant ISM which is supported against collapse by mechanical heating \citep[e.g.][]{Smercina18,French22}.
We conclude that despite a significant amount of ISM and molecular gas,  little star formation has taken place over the past $\sim100$~Myr outside of the tiny nuclear disks in Arp~220.

\subsection{The Star Formation History of Arp~220}

\begin{figure}[!ht]
	\centering
\includegraphics[width=12.0cm]{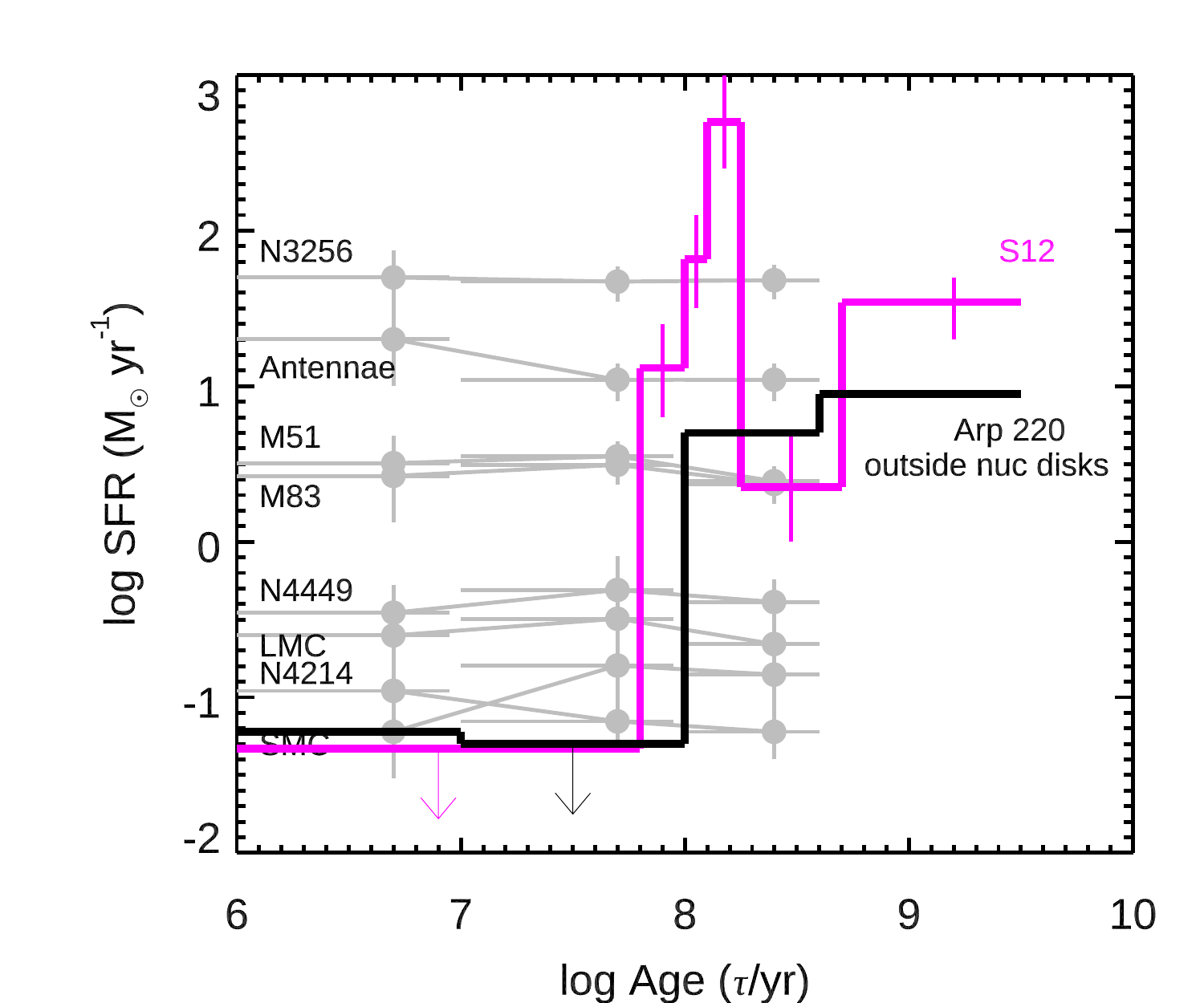}
	\caption{The recent star formation history of Arp~220 determined from its cluster population as described in Section~6 is shown as the black histogram. The cluster-based star formation history determined by \citet{Chandar21} for the post-starburst galaxy S12 is shown in pink. The histories determined for the eight labelled galaxies from integrated light measurements (not clusters) and presented in \citet{Chandar17} are shown for comparison, and are quite constant, unlike Arp~220 and S12. }\label{fig-compsfrs}
\end{figure}

\par 
The cluster-based star formation history of Arp~220 calculated in Section~6.2 is shown in Figure~\ref{fig-compsfrs}.  In the past $10$~Myr, Arp~220 has had a very modest SFR outside of the nuclear disks, estimated in the previous Section to be $\approx0.05~M_{\odot}~\mbox{yr}^{-1}$.

Because we did not find any clusters with ages in the $10-100$~Myr interval, we place an upper limit ($<0.05~M_{\odot}~\mbox{yr}^{-1}$) to the SFR over this period.
Many clusters in Arp~220 have ages of 100~Myr and older.  We estimate that the SFR over the $100-400$~Myr time period was between $\sim3-9~M_{\odot}~\mbox{yr}^{-1}$, similar to that found in many spiral galaxies.

The star formation history of Arp~220 tells us that around 100~Myr ago, star and cluster formation shut off nearly completely throughout 99.9\% of the area of Arp~220, i.e. everywhere but the nuclear disks and a few small localized nearby regions.  The ages and masses of the cluster population in Arp~220 suggest that star formation was rapidly and effectively quenched, quite different from the situation observed in all nearby spirals, irregulars, and dwarf starburst galaxies, but similar to what occurs in post-starbursts. 
\par 
Figure~\ref{fig-compsfrs} compares our estimated star formation history for Arp~220 with that of 8 well-known 'normal' star forming galaxies (shown in gray), including irregulars, dwarf starbursts, spirals, and on-going mergers.  All of these systems have had fairly constant rates of star formation over at least the past $\sim400$~Myr, the time frame over which we are able to estimate the SFR without worrying about the age/metallicity degeneracy.  Arp~220 by contrast, experienced a sharp drop in it's SFR about 100~Myr ago over its entire area outside of the tiny nuclear disks, where almost all star formation is currently concentrated.
The SFH of Arp~220 plotted in Figure~\ref{fig-compsfrs} is most similar to that of the post-starburst galaxy S12 \citep{Chandar21}, which had a similar sharp cutoff, with very little star formation over the last $\sim80$~Myr.  
Arp~220 didn't experience any major 'bursts' of SF (outside of the nuclear disks) as we typically think of in post-starbursts, and in fact was forming stars at a fairly modest $\approx3-9~M_{\odot}~\mbox{yr}^{-1}$.  This is very different from the short, extremely intense burst of star formation experienced by the post-starburst galaxy S12, right before star formation shut off.\footnote{The color-color diagram and resulting age-mass distribution of clusters in S12 are unusually bursty, including when compared with Arp~220.  The narrow age bins shown for S12 in Figure~\ref{fig-compsfrs} are driven by that particular system's unusual cluster population; see \citet{Chandar21} for details.} It is somewhat mysterious why star formation shut off $\sim100$~Myr ago throughout almost all of Arp~220. 

The tidal features on the west side of Arp~220 all have similar ages of $\approx1-2$~Gyr, while those on the east side are a bit younger, with estimated ages somewhere between $600-800$~Myr. This finding indicates that Arp~220 may have experienced multiple interactions or mergers in the past. 
The ages of the clusters in the tidal tails indicate that the events that created these features  pre-date the event that shut off star formation around 100~Myr ago. 
The estimated ages and locations of the clusters found here can help constrain future simulations of the merger history of Arp~220.

\subsection{Arp~220: A Shocked, Post-starburst System (SPOG)? }

The high infrared luminosity of Arp~220, its two prominent nuclear disks, dusty central region, and numerous tidal tails are all consistent with the expectations of an evolving merger. Our analysis of the cluster population confirms that the main body of the system is now essentially in a post-starburst phase, with almost all current star formation confined to the two $\sim150$~pc nuclear disks.\footnote{While Arp~220 is considered the archetypal ULIRG in the modern universe, it is not typical of IR-bright galaxies in the early universe, where AGN were being actively fueled by large amounts of gas. }  
\par 
Arp~220 therefore, appears to be in the process of transforming from a blue, star-forming galaxy, to a red, quiescent galaxy. Mergers and interactions are known to transform disk galaxies into spheroidal ones, to drive large amounts of gas into the nuclear regions, and to trigger star formation and the growth of central supermassive black holes.  This activity can trigger feedback from the AGN \citep[e.g.][]{DiMatteo05,Feruglio10,Cicone14,Cicone15}, and quench star formation, as seen in Arp~220.
\par 
 Arp~220 is not typically classified as a post-starburst galaxy, despite its dominant intermediate-age stellar population, because its integrated spectrum shows strong H$\alpha$ emission.  Nearly all of this emission however, arises from shocked gas driven by AGN activity, rather than from recent star formation.  
 Almost all current star formation is concentrated into the very small nuclear disk areas.
 These properties indicate that Arp~220 may be a 'Shocked POststarburst Galaxy' or a 'SPOG'.  
 \par 
 Shocked Post-starbursts are at an earlier phase in the galaxy transformation process than traditional post-starbursts. 
Post-starbursts were initially defined to be galaxies with A-type stellar spectra and {\em no} line emission.  Traditional searches however, are biased against galaxies which host AGN \citep[e.g.][]{Wild09,Kocevski11,Cales11,Cales13,Alatalo14a} because AGN can power significant H$\alpha$ and [OII] emission. 
Systems like NGC~1266, for example, would not be included in a traditional post-starburst search because it has shock-powered H$\alpha$ emission driven by an AGN \citep[e.g.][]{Davis12,Alatalo14b}, similar to Arp~220. 
In order to identify and study these types of galaxies, the Shocked POst-starbust Galaxy Survey (SPOGS), catalogued galaxies in a post-starburst phase which also have nebular lines excited by shocks instead of by star formation \citep{SPOGS}.
 It is possible that many transitioning galaxies go through a SPOG phase. 
 The star-formation in Arp~220 has been mostly quenched (outside of the AGN-hosting nuclear-disks), and nearly all of the H$\alpha$ emission arises from AGN-driven shocked gas.  Arp~220 also appears to have experienced recent merger events, and to be transitioning to a post-starburst phase, consistent with the features of a SPOG.

\section{Summary and Conclusions}\label{sec:summary}
Using new and archival multi-band $HST$ imaging taken as part of the CCD\&G survey, we have detected and studied the stellar clusters formed in Arp~220, the closest ULIRG.  The observations cover the near-ultraviolet through the near-infrared in 7 broad-band filters (NUV, U, B, V, I, J, \& H) and also include two narrow-band filters (H$\alpha$ and P$\beta$).
The high resolution images give one of the clearest views of the structures, gas, dust, and stellar clusters in Arp~220 to date, although they cannot resolve structure in either of the two very compact, nuclear disks.

We find that unlike the situation in actively star-forming galaxies, the dominant H$\alpha$ and Pa$\beta$ emission arise from different locations within Arp~220, and are driven by different processes. Approximately 90\% of the H$\alpha$ emission is generated by a shock-driven bubble emanating from the AGN in the western nucleus.  Less than 1\% is measured from 4 very young ($\lea 6$~Myr), low mass ($\lea 2\times10^4~M_{\odot}$) clusters located a few arseconds away from the nuclei.  Half of the Pa$\beta$ emission is emitted from the $\sim150$~pc nuclear disk regions, presumably due to some combination of the AGN and extreme star forming activity, while the rest is in a diffuse, extended kpc-scale disk surrounding the nuclei.  Less than 1\% of the Pa$\beta$ emission comes from the 4 young clusters.  A careful search of the  near-infrared images do not reveal any optically obscured, recently formed massive clusters.

We estimated the range of reddening and extinction affecting clusters within the central $12.5\arcsec$,  and implement a flexible SED fitting technique which allows the maximum $A_V$ to vary based on the amount of reddening found in the region.
The ages, extinctions, and masses of clusters are estimated by fitting their measured photometry with stellar evolutionary model predictions from \citet{Bruzual03}. 

One key result is that we only detect four very young ($\tau < 10$~Myr), low mass ($M \lea 2\times10^4~M_{\odot}$) clusters in Arp~220, with no detections of embedded clusters in the P$\beta$ line map.  Our finding differs from previous age-mass estimates of clusters from \citet{Wilson06}, who suggested there may be a population of very massive ($M \gea 10^6~M_{\odot}$), very young ($\tau < 10$~Myr) clusters forming in Arp~220, because they made an incorrect assumption regarding cluster age and reddening, but had insufficient data to directly break the age-reddening degeneracy.  The full suite of broad- plus narrow-band observations now available shows that their (reasonable) assumption about reddening was incorrect.  We find no clusters with estimated ages between 10 and 100~Myr.  Outside of the nuclear disks, Arp~220 is essentially in a post-starburst phase, with very little current star formation.

While star formation halted abruptly approximately 100~Myr ago, it was fairly continuous earlier than that, up to a few Gyr ago.  We estimated the rate of star formation in the $<10$~Myr, $10-100$~Myr, $100-400$~Myr, and 400~Myr$-3$~Gyr age intervals from a calibration between the well-studied cluster masses and star formation rates in 8 nearby star-forming galaxies. 
We estimated a star formation rate of 
$\approx3-9~M_{\odot}~\mbox{yr}^{-1}$ in the age interval
$\approx100-400$~Myr, similar to that found in many nearby spiral galaxies, but not as high as found in many actively merging systems like the the Antennae and NGC~3256.  This modest rate and continuous rather than bursty star formation history is intriguing, because it is not fully clear what event caused the star formation to shut off in the main body of Arp~220 and drive the system into a post-starburst phase.
The interaction/merging event(s) that created tidal tails T1-T5 occurred well before the shutoff of star formation based on cluster ages, so are unlikely to be the culprit. Tidal feature T6 has an estimate age that is close to, although somewhat older than, the time star formation shut off. The strong shock-driven H$\alpha$ emission coupled with the dominant intermediate age stellar populations found in Arp~220 suggest that this system is similar to the class of  more distant galaxies known as Shocked Post Starburst Galaxies or SPOGs.

\acknowledgements
R.C. acknowledges support from NASA/HST grant GO-15469, and thanks Adam Smercina, Decker French, and the anonymous referee for very helpful discussions about post-starburst galaxies.
\bibliography{master}
\end{document}